\documentclass[acmsmall,screen]{acmart}
\usepackage{makecell} 
\usepackage{multirow} 
\usepackage{textcomp} 
\usepackage{pifont} 
\usepackage{xspace} 
\usepackage{appendix} 
\usepackage{framed} 

\newenvironment{takeaway}{
  \par\addvspace{0.3em}  
  \noindent\begin{minipage}{\linewidth}
  \begin{snugshade}
  \setlength{\leftskip}{0.3em}
  \setlength{\rightskip}{0.3em}
  \noindent
}{
  \end{snugshade}
  \end{minipage}
  \par\addvspace{0.3em}  
}

\newcounter{jsonlisting}

\newcommand{\my}{\mbox{\emph{SAP}}\xspace}

\definecolor{shadecolor}{gray}{0.9}

\setcopyright{cc}
\setcctype{by}
\acmJournal{TOSEM}
\acmYear{2026} \acmVolume{1} \acmNumber{1} \acmArticle{}
\acmMonth{1} \acmDOI{10.1145/3788692}

\begin{document}

\title{A Large Scale Empirical Analysis on the Adherence Gap between Standards and Tools in SBOM}


\author{Chengjie Wang}
\email{chengjie2021@iscas.ac.cn}
\orcid{0009-0008-4445-9709}
\authornote{Intelligent Software Research Center, Institute of Software, Chinese Academy of Sciences, Beijing, China}
\authornote{University of Chinese Academy of Sciences, Beijing, China}

\author{Jingzheng Wu}
\email{jingzheng08@iscas.ac.cn}
\orcid{0000-0001-5561-9829}
\authornotemark[1] 
\authornote{Key Laboratory of System Software (Chinese Academy of Sciences), Beijing, China}

\author{Hao Lyu}
\email{lyuhao23@mails.ucas.ac.cn}
\orcid{0009-0007-0580-3346}
\authornotemark[1] 
\authornotemark[2] 

\author{Xiang Ling}
\email{lingxiang@iscas.ac.cn}
\orcid{0000-0002-7377-7844}
\authornotemark[1] 
\authornotemark[3] 
\authornote{Xiang Ling is the corresponding author.} 

\author{Tianyue Luo}
\email{tianyue@iscas.ac.cn}
\orcid{0000-0001-7407-8255}
\authornotemark[1] 

\author{Yanjun Wu}
\email{yanjun@iscas.ac.cn}
\orcid{0000-0002-1823-0459}
\authornotemark[1] 
\authornotemark[3] 

\author{Chen Zhao}
\email{zhaochen@iscas.ac.cn}
\orcid{0009-0005-3386-0335}
\authornotemark[1] 



\begin{abstract}
A Software Bill of Materials (SBOM) is a machine-readable artifact that systematically organizes software information, enhancing supply chain transparency and security. To facilitate the exchange and utilization of SBOMs, organizations such as the Linux Foundation and OWASP have proposed SBOM standards. Following standards, organizations have developed tools for generating and utilizing SBOMs. However, limited research has examined the adherence of these SBOM tools to standard specifications, a gap that could lead to compliance failures and disruptions in SBOM utilization.
This paper presents the first large-scale, two-stage empirical analysis of the adherence gap, using our automated evaluation framework, \my. The evaluation, comprising a baseline evaluation and a one-year longitudinal follow-up, covers 55,444 SBOMs generated by six SBOM tools from 3,287 real-world repositories. Our analysis reveals persistent, fundamental limitations in current SBOM tools: (1) inadequate compliance support with policy requirements; (2) poor tool consistencies, including inter-tool consistency rates as low as 7.84\% to 12.77\% for package detection across languages, and significant longitudinal inconsistency, where tools show low consistency with their own prior versions; and (3) mediocre to poor accuracy for detailed software information, \textit{e.g.}, accuracy of package licenses below 20\%. We analyze the root causes of these gaps and provide practical solutions. All the code, replication docker image, evaluation results are open sourced at GitHub and Zenodo\footnote{Available at \url{https://github.com/dw763j/SAP} and \url{https://doi.org/10.5281/zenodo.14998624}.} for further researches.
\end{abstract}


\received{4 August 2025}
\received[revised]{19 October 2025}
\received[accepted]{31 December 2025.}


\maketitle

\section{Introduction}


Modern software development increasingly relies on software supply chains, the complex networks of third-party libraries, frameworks, and dependencies that underpin contemporary applications~\cite{synkshift}.
Code reuse facilitated by software supply chains substantially improves development efficiency in modern software projects \cite{ntiaframing}. 
However, supply chain opacity may lead developers to inadvertently incorporate vulnerable or license-violating software~\cite{log4j,alkhadra2021solar,sonatypesotsc,przymus2025wolves}.
These risks necessitate proactive governance to address security and compliance challenges~\cite{zahan2023software, 10174240, 10315780, 10.1145/3640824.3640866, 10176156}.

A Software Bill of Materials (SBOM) is a structured document detailing software metadata across multiple data fields, aiming to enhance software transparency and security~\cite{o2024assessing, 10.1145/3605770.3625207, trivy, camp2021sbom}.
Consequently, their adoption is widespread. For instance, the Linux Foundation conducts a survey that reports that 90\% of the 341 surveyed organizations are implementing or planning SBOM strategies~\cite{readiness2022}. Furthermore, governments in the US and EU have initiated policies that leverage SBOMs for improved security management on the software supply chain~\cite{executiveorder2021,eu}.

To ensure the software supply chain transparency and security, organizations propose SBOM standards to standardize the process of exchanging software metadata. These standards provide an interoperable and consistent data structure for the SBOM, enabling effective exchange and processing. Among the community, key SBOM standards include the Software Package Data Exchange (SPDX) proposed by the Linux Foundation, the CycloneDX by the OWASP, and Software Identification Tagging (SWID Tagging) by the National Institute of Standards and Technology (NIST)~\cite{spdxspec, cdx, swidtag, spdxintro, cdx-one-paper}.
Guided by these standards, organizations develop SBOM tools to generate and consume SBOMs~\cite{cdxgen, gh-sbom, ort, sbom-tool, scancode,syft}. They enable the automated generation and consumption of SBOMs, thereby promoting broader SBOM adoption~\cite{10336262, 10305922, tobar2025software, nocera2025adoption}.

However, for SBOMs to fulfill their purpose as interchangeable data exchange artifacts, the tools creating them must adhere to the SBOM standards. Existing studies already highlight that current tools are often insufficient, producing inaccurate or incomplete SBOMs~\cite{10.1145/3597503.3623347, mirakhorli2024landscape, 10.1145/3654442, 10.1109/ICSE48619.2023.00219}. From the perspective of the implementation process from SBOM standards to SBOM tools, these shortcomings could come from a deeper, more fundamental problem, which we term the \textit{``adherence gap''}.

The \textit{``adherence gap''} is the discrepancy between the required structure, content, and quality attributes defined by SBOM standards and the actual output generated by SBOM tools. When tools fail to fully adhere to these specifications, the resulting SBOMs suffer from compliance failures and inconsistencies, disrupting their utilization and hindering the goal of improving software supply chain transparency and security.

\begin{table}[t]\small
    \centering
    \caption{Inherent data quality attributes defined by ISO/IEC 25012~\cite{25012}}
    \label{tab:prin}
    \resizebox{\linewidth}{!}{
    \begin{tabular}{p{0.1\linewidth}p{0.53\linewidth}p{0.3\linewidth}}
        \toprule
        \textbf{Attribute} & \textbf{Definition} & \textbf{Interpretation}\\
        \midrule
        \textit{Compliance} & The degree to which data has attributes that adhere to standards, conventions, or regulations in force and similar rules relating to data quality in a specific context of use. & SBOM tools adhere to the structure and mandatory data fields defined by the SBOM standards.\\
        \midrule
        \textit{Consistency} & The degree to which data has attributes that are free from contradiction and are coherent with other data in a specific context of use. & SBOM tools produce consistent information in SBOMs between each other.\\
        \midrule
        \textit{Accuracy} & The degree to which data has attributes that correctly represent the true value of the intended attribute of a concept or event in a specific context of use. & SBOM tools produce accurate information about software.\\
        \bottomrule
    \end{tabular}
    }
\end{table}

Thus, we set out to evaluate this adherence gap between SBOM standards and tools. To achieve this, we focus on the fundamental function of SBOMs as data exchange artifacts. SBOMs are designed to be reliably shared and processed across organizational boundaries. As such, the SBOM tools must adhere to the SBOM standards with the lens of established data quality principles, \textit{e.g.}, characteristics from ISO/IEC 25012~\cite{25012}. Table~\ref{tab:prin} presents the three data quality attributes that we systematically analyse, complying with ISO/IEC 25012: compliance, consistency, and accuracy. 
These attributes address distinct but interdependent quality dimensions that form a hierarchical dependency chain. Compliance establishes the foundation: without structural adherence to standards, SBOMs cannot be reliably processed. Consistency builds upon this foundation to enable SBOM interchangeability across tools. Accuracy represents the highest utility layer, determining whether SBOMs can effectively inform security and compliance decisions.

We design \my, an automated and extensible \underline{S}BOM g\underline{ap} evaluation framework to assess these attributes between SBOM standards and SBOM tools. It mainly consists of three modules:
first, \my executes SBOM tools through a standardized workflow in the generation module to generate SBOMs (\S \ref{sec:generation_module}).
It then draws information from SBOMs in the extraction module (\S \ref{sec:extract_module}) and assesses the data quality attributes between SBOM standards and SBOM tools in the evaluation module~(\S \ref{sec:evaluation_module}).

For this study, we employ a two-stage design to provide a dynamic view of the SBOM ecosystem. We first collect 3,287 real-world GitHub repositories spanning C/C++, Java, and Python languages (\S \ref{sec:drepo}). 
We conduct a baseline evaluation using six advanced SBOM tools at October 2024, and then perform a one-year longitudinal follow-up with their October 2025 versions, yielding a total of 27,795 and 27,649 SBOMs, respectively. The follow-up evaluations are commonly utilized for measuring evolving targets, which is suited for the evolving SBOM tools~\cite{mckenzie2012beyond, roestenberg2011long}.
Furthermore, we establish a ground truth dataset to evaluate the accuracy of SBOMs (\S \ref{sec:gt}).

We evaluate these baseline SBOMs with \my, in which 26,186 (94.2\% out of all the 27,795 baseline SBOMs) are successfully processed.
We identify three major gaps from the evaluation results:
\begin{itemize}
    \item \textbf{Inadequate policy compliance:} SBOM tools cannot fully satisfy key government-mandated requirements, such as the minimum SBOM elements requirements from the NTIA, posing risks to policy adherence.
    \item \textbf{Poor inter-tool consistency:} We observe inconsistencies across tools, averaging between 7.84\% and 12.77\% on package detection for different languages, in both data format and content. Moreover, we observe that SBOM tools exhibit inconsistencies when generating SBOMs across different standards, which undermines SBOM compatibility and interchangeability.
    \item \textbf{Mediocre to poor accuracy:} Tools exhibit mediocre accuracy in package detection and poor accuracy for specific fields within packages, critically impairing their utility for crucial use cases like dependency management and license compliance.
\end{itemize}
Furthermore, our longitudinal follow-up analysis reveals a fourth critical finding:
\begin{itemize}
    \item \textbf{Persistent gaps and ecosystem volatility:} Our one-year follow-up analysis reveals that while individual tools kept evolving, the fundamental adherence gaps persist. This highlights the systemic and deep-rooted nature of these challenges and exposes the ecosystem's overall volatility, proving that tool evolution itself is a significant source of inconsistency.
\end{itemize}

We investigate the root causes of these gaps, attributing them to unclear standard constraints and validation rules, and ambiguous scope definitions within both SBOM standards and tools.

In summary, this paper makes the following contributions:

\begin{itemize}
    \item We conduct the first large-scale longitudinal empirical analysis of the adherence gap between SBOM standards and SBOM tools, using multilingual real-world repositories.
    \item We propose an automated and extensible SBOM gap evaluation framework, \my. It enables the broader research community to systematically assess the adherence of SBOM tools to standards in diverse scenarios.
    \item We identify major gaps in compliance, consistency, and accuracy, and further demonstrate their persistence and volatility through a one-year longitudinal study. Our findings reveal that these gaps are systemic and that the ecosystem's evolution is unpredictable, posing ongoing risks to software supply chain management. We analyze the root causes of these gaps and provide practical solutions. We open source all the replication code and results at GitHub~\cite{github-replication} and Zenodo~\cite{2025_14998625} for further researches.
\end{itemize}

\section{Background}

\subsection{SBOMs and SBOM standards}
\label{sbom-standards}

SBOM standards define the necessary structure and data fields to facilitate the effective exchange and utilization of SBOMs~\cite{spdxintro, cdx-one-paper}.
Of the three prominent standards, SPDX and CycloneDX are the most widely adopted~\cite{sbomhistory}.
In contrast, SWID tags serve as identifiers for software components and are not widely used as a separate SBOM document within the SBOM community~\cite{fossasbomformat}.
These standards aim for both interoperability (ensuring consistent interpretation of data) and interchangeability (enabling the exchange of SBOMs from different sources) of SBOMs, thus providing transparent and exchangeable information of the software supply chain. 
These features are common objectives in many document and data exchange standards, such as HTML and PDF~\cite{html, pdf}.

\begin{figure}[b]
  \centering
  \refstepcounter{jsonlisting}  
  \includegraphics[width=\linewidth]{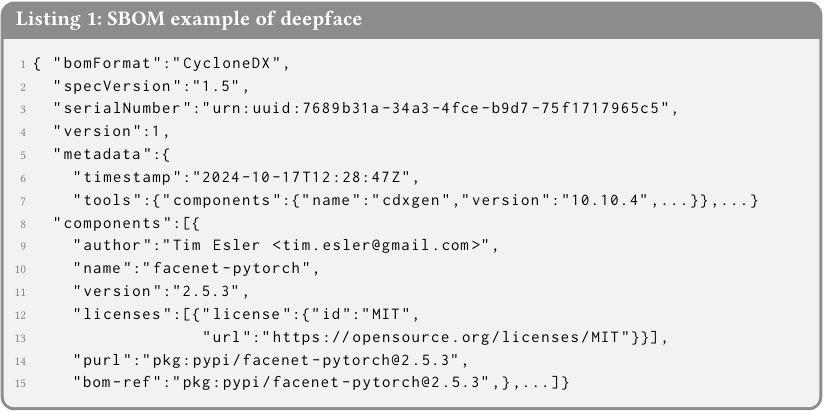}
  \Description{SBOM example of deepface.}
  \label{lst:sbom}
  \vspace{-2em}
\end{figure}

According to the standards, SBOMs typically include three main sections: the \textit{standard schema}, the \textit{metadata} of the SBOM, and the \textit{components} or \textit{packages}. Listing \ref{lst:sbom} provides a CycloneDX \texttt{json} format snippet of an SBOM for the deepface project~\cite{deepface} from Serengil, produced by cdxgen~\cite{cdxgen}.
The standard schema specifies which standard the SBOM complies with, \textit{i.e.}, CycloneDX 1.5, while the metadata includes detailed information about the SBOM, such as the creation timestamp and the creation tool.
The components section describes software dependencies, often as nested trees, with data fields detailing each component. For example, as shown in the snippet, the deepface includes a component named ``facenet-pytorch'' with a version of ``2.5.3'' and a license of ``MIT''.

\subsection{SBOM tools and SBOM standards}

While SBOM tools are instrumental in operationalizing SBOM standards~\cite{synkshift, sonatypesotsc}, their practical application is challenged by the ``adherence gap'', the discrepancy between the specifications defined by standards and the actual output generated by the tools. This gap may arise from two aspects: implementation and ongoing evolution.

\textbf{Firstly}, an adherence gap can arise during the implementation of the SBOM tool. SBOM standards define comprehensive and flexible characteristics of SBOMs to guide the community. This flexibility can lead to divergent interpretations among developers, resulting in tools that do not fully adhere to the required structure and content from the outset. This commonly exists in the software development process~\cite{garlan1995architectural, devanbu2000software}, which leads to degraded capability of the software. In the case of SBOM, the generated SBOMs by tools may suffer from compliance failures before they are even utilized. \textbf{Secondly}, the dynamic and asynchronous nature of the SBOM ecosystem can lead to a potential adherence gap. Both SBOM standards and the tools themselves are subject to continuous and independent updates~\cite{cdxgen, ort, syft, spdx3.0}. This continuous evolution can introduce or widen the adherence gap over time, as in the software development process~\cite{lloyd2011evolutionary, lehman2003software, knuppel2017there}. While in the area of SBOM, it may lead to inaccurate and incomplete SBOMs.

Despite the critical importance of tool adherence for ensuring SBOM reliability, this issue remains largely underexplored. Prior studies on SBOM quality have often been conducted without a detailed scrutiny against standard requirements~\cite{fossa-blog, sbombenchmark, 10315783, xiao2025jbomaudit}, focused on package-level analysis with limited scope~\cite{cofano2024sbom, 10.1145/3605098.3635927}, or utilized synthetic projects that do not reflect the complexities of real-world software~\cite{halbritter2024accuracy, yu2024correctness}. Therefore, this paper aims to address this research gap by conducting the first large-scale empirical analysis of the adherence gap between SBOM standards and tools.

\section{Framework}

\begin{figure}[t]
  \vspace{-1em}
  \centering
  \includegraphics[width=0.8\linewidth]{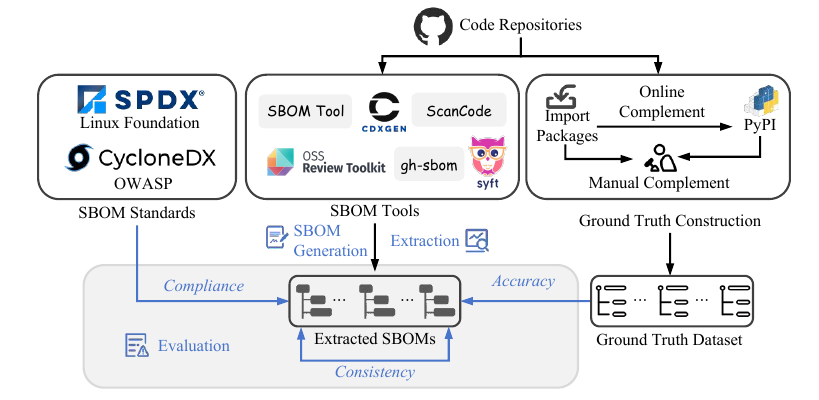}
  \caption{Overview pipeline of the empirical analysis based on \my.}
  \Description[Overview pipeline of the empirical analysis based on \my.]{The pipeline shows the process of SBOM extraction and evaluation. It includes SBOM standards (SPDX, CycloneDX), various SBOM tools extracting data from repositories, a ground truth construction process involving PyPI and manual efforts, and a final evaluation stage measuring compliance, accuracy, and consistency.}
  \label{fig:overview}
\end{figure}

To systematically investigate the adherence gap between SBOM standards and tools, we design and implement \my, an automated evaluation framework. The design of \my follows a logical three-stage pipeline to ensure a reproducible and comprehensive analysis, as illustrated in Figure~\ref{fig:overview}.

The pipeline begins with large-scale data acquisition, where the \textit{SBOM generation module} systematically invokes various SBOM tools to generate a diverse corpus of SBOMs from real-world software projects (\S \ref{sec:generation_module}). Subsequently, to enable a direct and uniform comparison, the \textit{SBOM extraction module} parses these heterogeneous SBOMs, normalizing their structure and content into a consistent format (\S \ref{sec:extract_module}). Finally, the \textit{Evaluation module} serves as the analytical core, applying a suite of metrics to the normalized data to quantitatively assess the adherence gap across key quality attributes (\S \ref{sec:evaluation_module}). The remainder of this section details the design of each module.

\subsection{SBOM generation module}
\label{sec:generation_module}
The primary objective of the SBOM generation module is to systematically construct a large-scale, diverse, and reproducible corpus of SBOMs for our empirical analysis. To achieve this, the module is designed around a three-stage process: (1) collecting a representative dataset of real-world software repositories, (2) establishing rigorous criteria to select advanced SBOM tools for evaluation, and (3) implementing an automated pipeline for the consistent generation of SBOMs. The following subsections detail each of these stages.

\subsubsection{Software collection}

\label{sec:software-collection}

The foundation of our empirical analysis is a dataset of software projects that is both representative of real-world development practices and suitable for automated evaluation.
To construct this dataset, we design a collection pipeline that sources repositories from GitHub. The selection process begins by querying for repositories using language-specific labels and filtering out forks to ensure project uniqueness and relevance. We use the number of stars as a proxy for project popularity and quality, collecting repositories in descending order of stars to prioritize influential projects. The collection is performed using \texttt{git clone}, which preserves the version control history that SBOM tools may rely on for their analysis. 
This structured approach yields a curated, high-quality dataset of popular software projects, organized by programming language to facilitate the subsequent stages of tool evaluation.

\subsubsection{SBOM tool selection criteria}

\label{sec:sbom-tool-select}

The selection of SBOM tools for our evaluation is guided by a set of rigorous criteria designed to ensure our analysis is relevant, reproducible, and generalizable.

We require that candidate tools satisfy the following five criteria:

\begin{itemize}
    \item \textit{\textbf{Official endorsement}}: Prioritizing tools officially listed or recommended by SBOM standard communities (\textit{e.g.}, SPDX, CycloneDX~\cite{spdx-tools, cdx-tools}), which indicates foundational standard compliance and a higher probability of quality.
    \item \textit{\textbf{Open source and accessibility}}: Tools should be open-source and publicly accessible to foster reproducibility and wider community adoption.
    \item \textit{\textbf{Multi-language capability}}: Tools should support SBOM generation across diverse programming language ecosystems to ensure the generalizability of evaluation results.
    \item \textit{\textbf{CLI automation}}: Tools should provide a command-line interface (CLI) for direct software analysis and the automated generation of SBOMs in standard formats.
    \item \textit{\textbf{Active maintenance}}: Tools should demonstrate continuous maintenance and an active support community to ensure relevance and the exclusion of obsolete projects.
\end{itemize}

These criteria not only evolve common metrics of existing studies, such as open source, accessible, and actively maintained tools~\cite{cofano2024sbom,halbritter2024accuracy, yu2024correctness}, but also further extend the recommended tools from standard community and multilingual tools to conduct the gap analysis between SBOM standards and tools in multilingual software.

\subsubsection{SBOM generation}

\label{sec:sbom-generation}

With the software repositories and tools in place, the final stage of this module is the automated generation of SBOMs. The primary objectives of our generation pipeline are to ensure consistency across all tools and to guarantee the reproducibility of our results.

To achieve this, we design a standardized execution protocol. First, to mitigate environmental dependencies and ensure reproducibility, each SBOM tool is installed and executed within the Docker container according to its official documentation. Second, we apply a consistent configuration scheme for each tool, providing the local file directory of a repository as input and activating a comprehensive set of analysis options to maximize the depth of the scan, \textit{e.g.} the target SBOM standard, compatible analysis types, and output format. The output is uniformly configured to the JSON format of the tool's supported standard, which streamlines the subsequent extraction process.

This controlled process results in a large corpus of SBOMs generated in a consistent and methodologically sound manner.

\subsection{Extraction module}
\label{sec:extract_module}

The SBOMs generated in the previous stage are heterogeneous, originating from different tools and conforming to various standard schemas, \textit{e.g.}, SPDX, CycloneDX. To enable a systematic and uniform comparison, the extraction module serves as a critical normalization layer. Its primary objective is to parse these diverse SBOMs and transform them into a single, consistent, and structured data representation for subsequent evaluation.

The module implements a multi-step normalization pipeline. The process begins with an initial validation phase, where each input SBOM is checked for well-formed JSON and structural conformance to its declared standard. Upon successful validation, the pipeline identifies the SBOM's schema, \textit{e.g.}, CycloneDX, to invoke the corresponding standard-specific parsing logic. The parser then systematically traverses the document's hierarchy. It first extracts and flattens top-level metadata into a simple key-value format. Subsequently, it iterates through each software package, extracting a comprehensive set of data fields associated with it. To ensure structural consistency in the final output, any field that is absent in the original SBOM is populated with a predefined placeholder, \textit{e.g.}, ``NE'' for ``Not Exists''.

Upon completion of this process, the extracted data from each SBOM is serialized into a unified JSON object with our designed consistent schema, represented as: \texttt{``SBOM\ metadata;\ [packages\  \{field\_a: value, field\_b:value ...\};\ ...]''}.
Take the \textit{license} as an example, the module extracts it from a component\footnote{~In CycloneDX, ``component'' corresponds to ``package'' in SPDX.}, following the path: SBOM (JSON)→components→component →licenses →license, as defined in the CycloneDX standard.
By systematically transforming a corpus of heterogeneous SBOMs into a uniform and structured dataset, the extraction module provides the clean and reliable foundation necessary for the quantitative analysis performed by the evaluation~module.

\subsection{Evaluation module}
\label{sec:evaluation_module}

With a normalized dataset of SBOMs, the evaluation module quantitatively measures the adherence gap. Our evaluation methodology is structured around three fundamental data quality attributes derived from the ISO/IEC 25012 standard: compliance, consistency, and accuracy. Each attribute assesses a distinct, hierarchical dimension of the adherence gap. Compliance serves as the foundational layer, evaluating structural adherence to the standard. Consistency builds upon this to assess the interchangeability of SBOMs. Finally, accuracy represents the highest utility layer, measuring the correctness of the SBOM data against ground truth. The following subsections detail the specific metrics used to evaluate each attribute.

\subsubsection{Compliance}

In the context of data quality, compliance can be multifaceted, encompassing both \textit{structural compliance} (adherence to the schema) and \textit{semantic validity} (the semantic correctness of the data). Acknowledging this distinction, our evaluation of compliance specifically focuses on structural compliance. We define this as the degree to which an SBOM tool populates the data fields prescribed by the SBOM standard's schema. This metric serves as a foundational measure of adherence; an SBOM must be structurally well-formed before its semantic content can be meaningfully evaluated for accuracy.

Based on this definition, \my measures the structural compliance for a given data field as the rate of its presence across all packages within an SBOM. A field is considered present if it contains a value and is not the ``NE'' (Not Exists) placeholder introduced during extraction (\S\ref{sec:extract_module}). The overall structural compliance for a data field $\mathcal{F}$ generated by a tool is then calculated as the average of these rates across all generated SBOMs, as defined in Formula~(\ref{eqa:compliance}).
\begin{equation}
  \label{eqa:compliance}
    \mathtt{Comp_{\mathcal{F}}} = \frac{1}{|\mathcal{R}|} \sum_{\gamma \in \mathcal{R}} \frac{\sum_{p \in \mathcal{P}_\gamma} \textbf{1}_{p, \mathcal{F}}}{|\mathcal{P}_\gamma|}
\end{equation}
where $\mathcal{R}$ is the set of valid SBOMs, $\mathcal{P}_\gamma$ is the set of packages within the $\gamma\text{-}th$ SBOM, and $\textbf{1}_{p, \mathcal{F}}$ is an indicator function that is 1 if the field $\mathcal{F}$ is present in package $p$, and 0 otherwise.

\subsubsection{Consistency}
\label{sec:evaluation-cons}

Building upon structural compliance, we evaluate consistency, which, according to ISO/IEC 25012 in the Table \ref{tab:prin}, is the uniformity of SBOM data for the same software artifact produced by SBOM tools. Specifically, the ``SBOM tool'' in the consistency evaluation is defined in a broad sense; thus, different versions of the same SBOM tool are also eligible for evaluation.

High consistency is a prerequisite for the interchangeability of SBOMs, ensuring that documents from different tools can be reliably used in downstream security and compliance pipelines. To assess this, our methodology employs a hierarchical evaluation, analyzing consistency at two distinct levels of granularity: (1) the \textit{package level}, which measures the agreement on the set of identified software components, and (2) the \textit{data-field level}, which measures the agreement on the specific attributes of those components.

\texttt{Packages:} 
The initial challenge in assessing consistency is to accurately pair corresponding package entries across SBOMs generated by different tools. This task is non-trivial, as SBOMs lack a universally adopted unique identifier for packages. While prior studies have often relied on pairing packages by name and version~\cite{yu2024correctness,10.1145/3605098.3635927,10.1145/3672608.3707940}, this approach can lead to ambiguous matches, particularly when multiple components share identical names or version strings, which is also observed in the existing study as up to 23.76\% duplications according to SBOM tools~\cite{yu2024correctness}.

To address this limitation, we propose a more robust heuristic: a triple-factor best-matching algorithm. This algorithm leverages three complementary data points for identification: package \textit{name}, \textit{version}, and \textit{Package URL (purl)}~\cite{purl}. A composite matching score is computed for each potential package pair $(p_1, p_2)$ using Formula~(\ref{eqa:match-1}). A pair is confirmed if its score meets a predefined threshold $\tau$, as shown in Formula~(\ref{eqa:match-2}).
\begin{equation}
\label{eqa:match-1}
\text{\texttt{match\_score}}(p_1, p_2) = \alpha \cdot \mathrm{sim}_{\mathrm{name}}(p_1, p_2) + \beta \cdot \mathrm{sim}_{\mathrm{version}}(p_1, p_2) + \gamma \cdot \mathrm{sim}_{\mathrm{purl}}(p_1, p_2)
\end{equation}
\begin{equation}
\label{eqa:match-2}
\text{\texttt{paired}}(p_1) = \begin{cases} 
\underset{p_2 \in P^\mathcal{Y}_\gamma}{\mathrm{argmax}}\ {\texttt{match\_score}}(p_1, p_2), & \mathtt{if}\ \max({\texttt{match\_score}}) \geq \tau \\
\mathrm{unmatched}, & \mathtt{otherwise}
\end{cases}
\end{equation}
where $p_1 \in \mathcal{P}^\mathcal{X}_\gamma$ and $p_2 \in \mathcal{P}^\mathcal{Y}_\gamma$ are the packages scanned from software $\gamma$ by the tool $\mathcal{X}$ and $\mathcal{Y}$, the $\alpha$, $\beta$ and $\gamma$ are weights for different dimensions, which are default set to 1.0. The $\mathrm{sim}_{\mathrm{name}}$, $\mathrm{sim}_{\mathrm{version}}$, and $\mathrm{sim}_{\mathrm{purl}}$ are the metrics for each dimension, whose details are described in the metrics for data fields. The $\tau$ is the threshold of the minimum matching to ensure true matches, which is set to 2.0 in our evaluations. Specifically, \my requires the match of the package name as the foundation; if the package name is not matched, the package will not be matched either.

The proposed triple-factor best-matching algorithm ensures a consistent and reproducible basis for the package pairing process.
Once package pairs are identified, the package-level consistency ($\mathtt{Cons_{\mathcal{P}}^{\mathcal{X}\leftrightarrow\mathcal{Y}}}$) between SBOMs from tool $\mathcal{X}$ and $\mathcal{Y}$ is calculated using Formula~(\ref{eqa:consistency-pkg}):
\begin{equation}
  \label{eqa:consistency-pkg}
  \mathtt{Cons_{\mathcal{P}}^{\mathcal{X}\leftrightarrow\mathcal{Y}}} = \frac{1}{|\mathcal{R}_c|} \sum_{\gamma \in \mathcal{R}_c} \frac{|\mathrm{Paired}_{\gamma}^{\mathcal{X}\leftrightarrow\mathcal{Y}}|} {\max(|\mathcal{P}_{\gamma}^{\mathcal{X}}|, |\mathcal{P}_{\gamma}^{\mathcal{Y}}|)}
\end{equation}
where $\mathcal{R}_c$ is the set of repositories commonly and successfully analyzed by both SBOM tools. For each repository $\gamma \in \mathcal{R}_c$, $\mathcal{P}_{\gamma}^{\mathcal{X}}$ and $\mathcal{P}_{\gamma}^{\mathcal{Y}}$ are the sets of packages scanned by tool $\mathcal{X}$ and tool $\mathcal{Y}$, respectively.
$\mathrm{Paired}_{\gamma}^{\mathcal{X}\leftrightarrow\mathcal{Y}}$ represents the set of successfully paired packages between SBOMs for repository $\gamma$.

Our designed triple-factor best-matching algorithm can distinguish the most consistent package of the multiple packages in the SBOM more accurately, and thus can provide more effective evaluation results compared to existing studies.

\texttt{Data fields within paired packages:} 
Within the matched package pairs, the consistency of the internal data fields is assessed using approaches tailored to the semantic and structural properties of each field. \my categorizes fields and applies illustrative evaluation strategies as~follows:

\textbf{(1) Exact value specifiers}, \textit{e.g.}, \textit{download URLs}, \textit{license}: 
These locators and enumerated values require strict identity; thus, an exact string match after normalization is needed.
Normalization includes breaking down values to atomize one for comparing, such as the \textit{license} values are broken into a canonical form, \textit{e.g.}, ``MIT AND Apache-2.0'' is parsed into ``MIT'' and ``Apache-2.0'', and then computing the match results.

\textbf{(2) Structured data}, \textit{e.g.}, \textit{package version}, \textit{purl}: 
These fields adhere to a specific format or specification, possessing an internal, parsable structure for comparing segment by segment.
For instance, \textit{package version}, when versions adhere to Semantic Versioning (SemVer) \cite{semver}, as recommended by CycloneDX, \my~employs a structured comparison: versions are parsed using the official ``semver'' library and compared segment by segment. 
The overall similarity score is a weighted sum of similarities for corresponding segments as detailed in Formula~(\ref{eqa:ver}). 
If a version string does not conform to SemVer, \my~defaults to the unstructured text (see next category).
\begin{equation}
    \label{eqa:ver}
     \mathtt{Cons_{\mathrm{SemVer}}}(v_1, v_2) = \sum_{i=1}^{N} \textbf{1} \cdot \lambda_i \cdot (1 - \frac{|n_i^1 - n_i^2|}{\max(n_i^1, n_i^2)})
\end{equation}
where $v_1, v_2$ are the two version strings being compared; $n_i^1$ and $n_i^2$ are the numerical values in stage $i$, $\lambda_i$ is weight for each of the $N$ stages (\textit{e.g.}, default values of 0.7, 0.2 and 0.1); and $\textbf{1}$ is an indicator that turns to 0 when a mismatch occurs in any previous stage.
Similarly, for \textit{purl}, \my adopts the official ``package-url parser'' and compares segment by segment.

\textbf{(3) Unstructured free-form text:} These fields contain natural language text or a string without a strict, machine-parsable syntax, such as \textit{copyright} statements or component descriptions. Minor variations in phrasing, typos, or formatting are common and may not signify a semantic difference. Consistency often implies high textual similarity rather than absolute character-for-character identity.
Jaro-Winkler similarity score~\cite{jaro1989advances} is well-suited for short texts like \textit{copyright} statements, as it accounts for matching characters and transpositions.

The modular architecture of \my~supports future extensions by allowing the integration of new metric plug-ins for emerging field types or improved comparison algorithms.

\texttt{Consistency rate:} Based on these tailored methods, the consistency for a data field $\mathcal{F}$ within paired packages is calculated as the average similarity or match rate, as shown in Formula~(\ref{eqa:consistency-field}):
\begin{equation}
  \label{eqa:consistency-field}
  {\mathtt{Cons_{\mathcal{F}}^{\mathcal{X}\leftrightarrow\mathcal{Y}}} = \frac{1}{|\mathcal{R}_c|} \sum_{\gamma \in \mathcal{R}_c} \frac{\sum_{p \in \mathrm{Paired}_{\gamma}^{\mathcal{X}\leftrightarrow\mathcal{Y}}} \mathcal{M}_\mathcal{F}(\mathcal{F}_p^\mathcal{X}, \mathcal{F}_p^\mathcal{Y})} {|\mathrm{Paired}_{\gamma}^{\mathcal{X}\leftrightarrow\mathcal{Y}}|}}
\end{equation}
where $\mathcal{M}_\mathcal{F}$ is the evaluation metric chosen for field $\mathcal{F}$ (yielding, \textit{e.g.}, a binary 0/1 for exact matches, or a similarity score between 0 and 1). $\mathcal{F}_p^\mathcal{X}$ and $\mathcal{F}_p^\mathcal{Y}$ are the values of field $\mathcal{F}$ in package $p$ from the SBOMs generated by both tools. ${\mathrm{Paired}_{\gamma}^{\mathcal{X}\leftrightarrow\mathcal{Y}}}$ is the set of paired packages for repository $\gamma$.

\subsubsection{Accuracy}
\label{sec:acc-design}
The final metric that one SBOM can be utilized for is accuracy, which measures the degree to which data correctly represents the true state of the software. While compliance and consistency ensure that an SBOM is well-formed and interchangeable, accuracy determines whether its content is factually correct and thus trustworthy for critical software supply chain management. The evaluation of accuracy is performed by comparing the tool-generated SBOMs against a manually curated ground truth (GT) dataset. Following our hierarchical approach, we assess accuracy at two levels: the package level and the data-field level.

\texttt{Package:} This assessment evaluates a tool's ability to correctly identify the complete set of software components. We treat this as a set-based comparison between the packages reported by the tool and those listed in the ground truth. Consequently, we employ the standard information retrieval metrics of precision and recall to quantify performance.
\begin{equation}
  \label{eqa:pre}
  \mathtt{Precision = \frac{|TP|}{|TP| + |FP|}};\ \mathtt{Recall = \frac{|TP|}{|TP| + |FN|}}
\end{equation}
Here, True Positives (TP) represent the set of packages correctly identified by the tool that are also in the GT. False Positives (FP) are packages reported by the tool but absent from the GT. False Negatives (FN) are packages present in the GT but missed by the tool.

\texttt{Data fields within paired packages:} For the set of correctly identified packages (the TPs), we conduct a deeper analysis to evaluate the accuracy of their specific data fields. This is accomplished by comparing the values reported by the tool, \textit{e.g.}, for license or version,  against the corresponding authoritative values in the ground truth. The same tailored comparison functions used in the consistency analysis are applied here to determine if the reported data is correct. This two-level approach provides a comprehensive measure of a tool's accuracy, from its high-level dependency detection down to the fine-grained correctness of the component metadata.

\section{Experiment Setup}

\subsection{Software dataset}
\label{sec:drepo}

Using \my, we conduct adherence gap evaluations between SBOM standards and tools using a real-world software dataset. The selection of our software dataset is guided by the principle of maximizing the generalizability of our findings. To this end, we curated a dataset of real-world software repositories designed to be representative of the key challenges in SBOM adoption. Our selection strategy is therefore based on three core pillars: policy relevance, ecosystem diversity, and the multilingual nature of modern~software.

\begin{table}[b]
  \centering
  \includegraphics[width=0.6\linewidth]{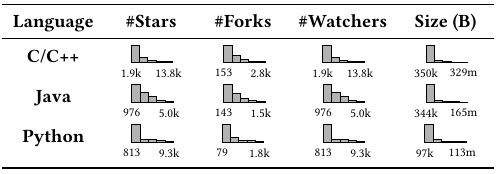}
  \caption{Statistical distribution of $D_{repo}$.}
  \label{tab:statistic_dataset}
  \vspace{-2em}
\end{table}

First, we selected C/C++, Java, and Python as the primary languages for our study due to their immense policy and industrial relevance. These languages are predominant in critical infrastructure and enterprise software sectors~\cite{stroustrup1999overview, javaanalysis, pythonindustries}, which are the explicit targets of emerging SBOM regulations such as the US Executive Order on Cybersecurity~\cite{executiveorder2021} and the EU Cyber Resilience Act~\cite{eu}. Second, this trio represents a broad spectrum of ecosystem diversity, presenting SBOM tools with fundamentally different challenges. This includes the minimal standardized packaging of C/C++, the mature and complex dependency trees in Java's Maven/Gradle ecosystem, and the flexible, script-based nature of Python. Evaluating tools against this diversity is crucial for assessing their robustness across different development paradigms.

For the data quality assurance, we constructed the dataset $D_{repo}$ by collecting 3,287 repositories, \textit{i.e.}, 1,030 C/C++ (combined), 1,098 Java, and 1,159 Python, from GitHub by the official RESTful API, each with at least 100 stars to filter out inactive or experimental projects. As shown in Table~\ref{tab:statistic_dataset}, the high popularity of these repositories ensures that our analysis is grounded in established software with significant community adoption, where SBOM accuracy is of practical consequence.

Moreover, while repositories have a primary language, we recognize that real-world projects are often multilingual. To account for this, we analyzed our dataset using GitHub's Linguist label~\cite{Linguist}. As shown in Table~\ref{tab:linguist-data}, our dataset, despite being primarily composed of C/C++, Java, and Python projects, also contains a significant volume of code in other languages such as Shell and JavaScript. In total, 310 languages occurred in the $D_{repo}$ dataset (nine listed in the table, 301 in the ``others'' column). This inherent multilingualism enriches our evaluation by ensuring that tools are tested against the complex, heterogeneous environments typical of modern software development.

This selection strategy directly addresses the limitations of prior studies that either used synthetic projects~\cite{cofano2024sbom} or focused on single-language ecosystems~\cite{10.1145/3605098.3635927, 10.1145/3672608.3707940}. While other important ecosystems, such as Rust, present unique challenges worthy of future investigation, expanding the primary analysis to include them constitutes mainly an engineering effort. We contend that our curated dataset is sufficiently representative to uncover the fundamental, cross-ecosystem adherence gaps that are the core focus of this paper.

\begin{table}[t]
\vspace{-1em}
\centering
\caption{Distribution of programming languages in the $D_{repo}$ dataset, as identified by the GitHub Linguist label. For clarity, only languages appearing in at least 500 repository-language entries are listed by name.}
\label{tab:linguist-data}
\resizebox{\linewidth}{!}{
  \begin{tabular}{ccccccccccc}
    \toprule
     \textbf{Language} & \textbf{Python} & \textbf{Shell} & \textbf{Java} & \textbf{C} & \textbf{Makefile} & \textbf{C++} & \textbf{HTML} & \textbf{JavaScript} & \textbf{Dockerfile} & \textbf{Others$^{\dag}$}\\
    \midrule
    \textbf{Repository Occurrences} & 1,787 & 1,533 & 1,237 & 1,176 & 1,036 & 736 & 715 & 590 & 511 & 6,899 \\
    \bottomrule
  \end{tabular}
  }
  
  \raggedright\footnotesize
  ~$^{\dag}$ An aggregate of 301 different languages, each with fewer than 500 repository occurrences.
\end{table}

\subsection{SBOM dataset}
\label{sec:dsbom}
\subsubsection{SBOM standard selection}
SPDX and CycloneDX are widely used SBOM standards when compared to SWID Tagging, as discussed in \S \ref{sbom-standards}.
SBOM tools typically support at least one of them~\cite{sok}, thus we select SPDX v2.3 and CycloneDX v1.5 as the target standards for our analysis. This selection ensures the generalizability of our evaluations.

\subsubsection{SBOM tool selection}

\label{sec:sbom-tool-real-select}
For the representativeness of analyses, we screen SBOM tools for evaluation using the criteria defined at \S \ref{sec:sbom-tool-select} as follows:
Initially, 302 tools are recommended by SBOM communities, of which 218 are open source.
Among them, 87 tools support analyzing software and building the SBOM, with 49 specifically designed for a single language.
Of the remaining 38 tools, 24 are libraries or plugins that cannot independently generate SBOMs via the command line CLI.
Eight tools have been inactive for at least two years or have fewer than 100 stars on GitHub.

After filtering, six tools remain: cdxgen@v10.10.4~\cite{cdxgen}, gh-sbom@v0.0.9~\cite{gh-sbom}, ort@v36.0.0-018~\cite{ort}, syft@v1.14.1~\cite{syft}, sbom-tool@v2.2.1~\cite{sbom-tool}, and scancode@v32.0.8~\cite{scancode}. 
Cdxgen is the official tool developed by the CycloneDX workgroup. 
Gh-sbom and sbom-tool are developed by reputable organizations, namely GitHub and Microsoft, respectively.
Ort, scancode, and syft are popular open-source SBOM tools in the community, each with thousands of stars on GitHub.
All these selected tools support the analysis of software in the three chosen programming languages.
The sbom-tool supports generating SBOMs in SPDX; cdxgen and scancode support CycloneDX; while gh-sbom, ort, and syft support both standards. The versions of the SBOM tools are the state as of October 2024, which form the baseline of our analysis.

\begin{table}[b]\small
  \caption{Statistics of generated SBOMs}
  \label{tab:statistics_sboms}
  \centering
  \begin{minipage}{0.8\linewidth}
  \centering
  \begin{tabular}{rrrrrrrr}
    \toprule
     & \textbf{cdxgen} & \textbf{sbom-tool$^{\dag}$} & \textbf{scancode} & \textbf{gh-sbom$^{\ddag}$} & \textbf{ort$^{\dag}$} & \textbf{syft$^{\ddag}$} & \textbf{Total}\\
    \midrule
    \textbf{SBOMs} & 3,243 & 3,287 & 3,275 & 6,314 & 5,102 & 6,574 & 27,795\\
    \bottomrule
  \end{tabular}

  \raggedright\footnotesize
  $^{\dag}$ Tool in SPDX, while the unmarked tools are in CycloneDX.\\
  $^{\ddag}$ Tools supporting both CycloneDX and SPDX.
  \end{minipage}
  \vspace{-1em}
\end{table}

\subsubsection{SBOM generation}
\label{sbom-generate}
Following the pipeline described in \my at \S \ref{sec:sbom-generation}, we install the selected SBOM tools and set up parameters for SBOM generation. 

Using all six SBOM tools, we analyze the 3,287 repositories in $D_{repo}$ and generate a total of 27,795 SBOMs. 
To promote open science, we release the SBOMs and related code on Zenodo\cite{2025_14998625}.
Table~\ref{tab:statistics_sboms}
shows the number of SBOMs generated by tools.
The sbom-tool and syft successfully generate SBOMs for every repository, which demonstrates the effectiveness of \my's orchestration.
The remaining tools failed to generate certain SBOMs due to their internal errors, \textit{e.g.}, the internal timeout encountered by ort.

\subsection{Ground truth dataset}
\label{sec:gt}

To rigorously evaluate SBOM accuracy, we need to create a ground truth dataset. Existing studies utilize synthesized projects~\cite{cofano2024sbom} or directly use metadata of language systems~\cite{xiao2025jbomaudit,10.1145/3672608.3707940,10.1145/3605098.3635927}. However, this may miss information that is not recorded by the developers in the metadata. Some GitHub repositories today may contain the SBOM itself, but it is also mainly generated by the SBOM tools~\cite{10336262}, thus cannot be utilized to evaluate SBOM tools.

To address these challenges, we construct $D_{gt}$ through a multi-stage verification process that combines automated extraction with human validation of real-world repositories. This approach ensures both ecological validity and measurement reliability.

\begin{table}[b]\small
  \caption{Statistics of the ground truth dataset $D_{gt}$.}
  \label{tab:bench}
  \centering
  \begin{tabular}{ccccc} 
    \toprule
     & \textbf{packages} & \textbf{supplier} & \textbf{version}  & \textbf{license}  \\
    \midrule
    \textbf{Repository-level} & 660 & 72 & 38 & 81 \\
    \textbf{Package-level}  &  - & 513 & 169 & 612\\
    \bottomrule
  \end{tabular}
  \vspace{-1em}
\end{table}

For our initial investigation into SBOM accuracy, we intentionally focused our ground truth creation on the Python ecosystem. This decision was driven by a balance of methodological necessity and representativeness. Firstly, Python's distinct characteristics, such as its explicit ``import'' statements and prevalent package managers like PyPI, enable a reliable and reproducible manual verification process. This is crucial for establishing a trustworthy baseline for our evaluation framework. Secondly, Python represents a critical and widely used ecosystem where SBOM accuracy is of paramount importance for security and compliance.

The $D_{gt}$ follows SPDX structure to maintain alignment with SBOM standards while containing only data fields verifiable through authoritative channels: \textit{package name}, \textit{version}, \textit{supplier}, and~\textit{license}.

Our construction process involves three verification stages:

\begin{itemize}
    \item \textbf{Metadata analysis}: We identify evident dependencies of both direct and transitive from standard metadata files such as ``requirements.txt'', ``setup.py'', and ``Poetry.lock'', also with substandard metadata files such as ``install.txt'' in the pip format
    \item \textbf{Code traversal}: Source code is analyzed by traversing all the import statements to capture dependencies missing from metadata, with standard library imports excluded
    \item \textbf{Authoritative validation}: Official APIs, including GitHub API and PyPI API, are queried for supplier and license information, with manual verification of ambiguous cases
\end{itemize}

To ensure high-fidelity results, two authors independently validate all entries, resolving discrepancies through discussion. This process specifically addresses the metadata inconsistencies observed in real-world repositories, \textit{e.g.}, non-standard file names or missing version information, that would confound purely automated approaches.

To balance rigor with feasibility, $D_{gt}$ comprises 100 randomly selected Python repositories under 10MB from $D_{repo}$, excluding toy projects, \textit{e.g.}, notebooks or single-file repositories. As Table \ref{tab:bench} shows, $D_{gt}$ contains 660 dependent packages with verified metadata, revealing critical real-world challenges: 74.4\% ((660-169)/660) of packages lack explicit version information, and only 77.2\% (513/660) packages have verifiable supplier information. These findings, unattainable through synthetic data or existing SBOMs, highlight the necessity of our verification approach for meaningful accuracy assessment.

This methodology directly addresses the limitations of prior work by providing ground truth that is both representative of real-world conditions and rigorously verified against authoritative sources, enabling valid assessment of SBOM tool accuracy where it matters most.

\section{Evaluation Results and Analyses}

This section presents the results and their analyses of the SBOM standard adherence gap evaluation conducted using \my. Following the data quality attributes defined in Table~\ref{tab:prin}, we formulate three Research Questions (RQs) in \S \ref{sec:rqs}. We report the results of the three RQs in \S \ref{sec:compliance-results}, \S \ref{sec:consistency-results}, and \S \ref{sec:accuracy-results}, respectively. Furthermore, we report the empirical analysis experiment results of the \my itself in \S \ref{sec:sap-empirical} to validate the effectiveness of \my.

\subsection{Research Questions}

\label{sec:rqs}

We detail the research questions that guide our evaluations as follows:

\begin{itemize}
    \item \textbf{RQ1 (Structural Compliance)}: To what extent do SBOM tools adhere to the mandatory data fields specified in SBOM standards and policy requirements? \textit{This foundational layer determines whether SBOMs can be reliably parsed and processed, a prerequisite for any downstream application. Without structural compliance, all higher-level SBOM functionality becomes impossible.}
    \item \textbf{RQ2 (Content Consistency)}: How consistent are SBOM tools in representing the same software components across different tools and standards? \textit{Building on structural compliance, consistency enables SBOM interchangeability, the ability to substitute SBOMs from different sources without modification. Without consistency, organizations cannot confidently exchange or rely on SBOM data across toolchains.}
    \item \textbf{RQ3 (Information Accuracy)}: How accurately do SBOM tools capture critical software metadata (\textit{e.g.}, license information, version) compared to ground truth? \textit{Accuracy determines practical utility. Without it, SBOMs cannot reliably inform security and compliance decisions, regardless of their structural compliance or consistency. This represents the ultimate measure of SBOM value realization.}
\end{itemize}

\subsection{RQ1. Compliance}

\label{sec:compliance-results}

\begin{table}[b]\small
  \centering
  \caption{Data field groups of the compliance evaluation.}
  \label{tab:compliance}
  \begin{tabular}{ll} 
    \toprule
    \textbf{Group} & \textbf{Data Fields}\\
    \midrule
    \textbf{Mandatory} & \makecell[l]{spec version, SBOM license, namespace, creator, bom format.}\\
    \midrule
    $\textbf{\textit{NTIA}}^+$ & \makecell[l]{creator, timestamp, package name, package version, supplier, \\ license, copyright, unique identifier, package relationships.}\\
    \bottomrule
  \end{tabular}
  \vspace{-1em}
\end{table}

\subsubsection{Evaluation settings}
We establish two groups of data fields as shown in Table~\ref{tab:compliance} for assessing compliance with both the structure of SBOM standards and governmental mandates.

For basic interoperability, SBOM standards define mandatory data fields.
SPDX defines four mandatory data fields (listed in the first four fields of ``\textbf{Mandatory}'' in Table~\ref{tab:compliance}), while CycloneDX mandates two: \textit{bom format} and \textit{spec version}. 

In addition to these standards' mandatory fields, government bodies and institutions require additional fields to comply with policy requirements.
The National Telecommunications and Information Administration (NTIA) defines minimum SBOM elements~\cite{minimum}, which the US government requires to ensure software security and transparency~\cite{minimum_publish}.
To enhance compliance evaluation, we extend the set of fields with additional institutional mandates (\textit{i.e.}, \textit{license} and \textit{copyright}), following existing works~\cite{10.1109/ICSE48619.2023.00219, 10.1145/3597503.3623347}.
We label this enhanced group as $\textbf{\textit{NTIA}}^\textbf{+}$ as shown in Table~\ref{tab:compliance}. 
Specifically, for the \textit{unique identifier} field, we adopt the trend of clearer information needs in NTIA, thus using the \textit{purl} in the evaluation~\cite{minimum}.

\begin{figure*}[t]
  \vspace{-1em}
  \centering
  \includegraphics[width=\linewidth]{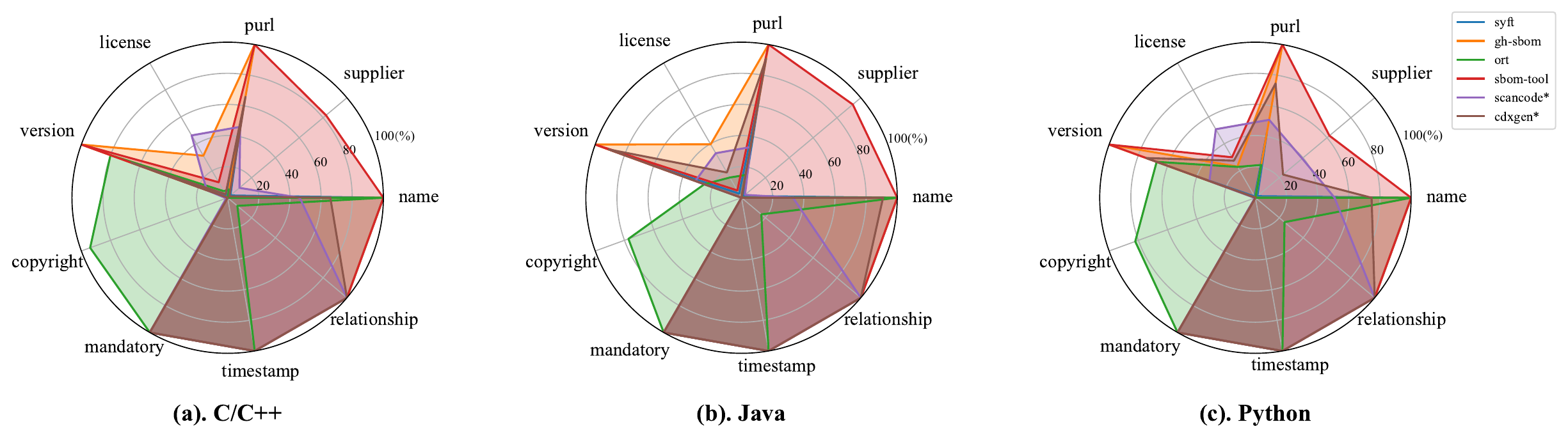}
  \caption{Compliance evaluation results across languages. The legend (top right) shows both CycloneDX and SPDX results, where the tools in CycloneDX are marked with~``*'' and tools in SPDX are unmarked. The ``mandatory'' axis includes results for all mandatory fields of each standard, while the remaining data fields are from $NTIA^+$ (the \textit{creator} field is overlapped in ``mandatory''). Best viewed in color.}
  \Description{Three radar charts compare the compliance of six SBOM tools across C/C++, Java, and Python. The dimensions include purl, supplier, name, relationship, timestamp, mandatory, copyright, version, and license. The charts show that while mandatory and basic fields like name/timestamp have high compliance, fields like copyright and purl vary significantly depending on the tool and the programming language.}
  \label{fig:compliance}
\end{figure*}

\subsubsection{Compliance results}

Figure \ref{fig:compliance} illustrates the understandable compliance evaluation results for all SBOM tools, covering the two groups of data fields detailed in Table~\ref{tab:compliance}. Moreover, the Table \ref{tab:compliance-data} provides the detailed compliance results numbers.
We report evaluation results for each tool based on the SBOM standard(s) it supports. 

Specifically, for tools supporting both CycloneDX and SPDX, \textit{i.e.}, gh-sbom, ort, and syft, we present results in SPDX. This choice is made because the SPDX standard, by default, records repository-level information in the same section as dependent packages, thereby ensuring the presence of at least one data field group for compliance evaluations.

\textbf{Mandatory fields:} All SBOM tools across the three languages can comply with the mandatory data fields of both SBOM standards, as shown in the ``mandatory'' column of Table~\ref{tab:compliance-data}, ensuring basic interoperability by the standards between SBOM files. However, besides these mandatory fields, more detailed fields fall short in poor compliance, as discussed below.

\textbf{$\textbf{\textit{NTIA}}^+$ policy-fields:} Compliance with the governmental requirements varies across tools and data fields.
The SBOMs generated by all tools generally adhere to the structure of their chosen standard.
However, for certain $NTIA^+$ fields, values are absent in SBOMs, leading to incomplete compliance as shown in Figure~\ref{fig:compliance}.
For example, sbom-tool achieves an average of 79.27\% compliance for the \textit{supplier} field across the three languages, while ort records 84.46\% on average for \textit{copyright}.
The other tools generally perform poorly in these two fields. This low compliance on SBOM tools can break the functionality of reliably parsing and processing SBOMs, which directly prevents the downstream applications that rely on these data fields.

\textbf{Across languages:} The compliance of SBOM tools is influenced by the programming language of the repository.
\textit{Package name}, \textit{package version}, and \textit{purl} show compliance variance across languages because these fields depend on language-specific characteristics.
For instance, ort achieves compliance scores of 80.07\%, 25.41\%, and 67.49\% for the \textit{package version} field in C/C++, Java, and Python repositories, respectively, highlighting this language-dependent impact.
Conversely, language-independent fields, including \textit{copyright} and \textit{license}, tend to show consistent results across languages. For instance, scancode achieves compliance of 46.28\%, 33.06\%, and 50.86\% for the \textit{license} across languages as shown in Table~\ref{tab:compliance-data}. Overall, the compliance of SBOM tools is highly affected by the languages of the software, highlighting the poor support of SBOM tools on multilingual software and the need for users to carefully select SBOM tools that fit the specific scenario.

\begin{table}[t]\small
    \vspace{-1em}
    \centering
    \caption{Detail compliance evaluation results across languages. The ``Mandatory'' column includes results for all mandatory fields of each standard, while the remaining data fields are from $NTIA^+$ (the \textit{creator} field is overlapped in ``mandatory''). All the results are reported in percentages.}
    \label{tab:compliance-data}
    \resizebox{\linewidth}{!}{
    \begin{tabular}{rrrrrrrrrr}
    \toprule
        & \textbf{Tool} & \textbf{Mandatory} & \textbf{name} & \textbf{supplier} & \textbf{purl} & \textbf{license} & \textbf{version} & \textbf{copyright} & \textbf{relationship} \\ \midrule
        \multirow{6}{*}{\rotatebox{90}{\textbf{C/C++}}} & syft & 100.00 & 100.00 & 2.28 & 48.19 & 0.99 & 48.95 & 0.00 & 100.00 \\ 
        & gh-sbom & 100.00 & 100.00 & 0.00 & 100.00 & 31.21 & 100.00 & 7.93 & 100.00 \\ 
        & ort & 100.00 & 100.00 & 0.00 & 5.48 & 4.87 & 80.07 & 93.97 & 8.10 \\ 
        & sbom-tool & 100.00 & 100.00 & 82.47 & 99.97 & 11.49 & 99.97 & 0.00 & 100.00 \\ 
        & scancode$^{\dag}$ & 100.00 & 46.28 & 9.89 & 46.28 & 46.28 & 14.33 & 1.02 & 100.00 \\ 
        & cdxgen$^{\dag}$ & 100.00 & 65.94 & 1.44 & 65.95 & 2.64 & 65.95 & 0.00 & 100.00 \\ 
        \midrule
        \multirow{6}{*}{\rotatebox{90}{\textbf{Java}}} & syft & 100.00 & 100.00 & 1.76 & 65.74 & 3.08 & 65.80 & 0.00 & 100.00 \\ 
        & gh-sbom & 100.00 & 100.00 & 0.00 & 100.00 & 39.65 & 100.00 & 9.10 & 100.00 \\ 
        & ort & 100.00 & 100.00 & 0.00 & 14.88 & 15.02 & 25.41 & 77.29 & 16.59 \\ 
        & sbom-tool & 100.00 & 100.00 & 93.27 & 100.00 & 5.43 & 100.00 & 0.00 & 100.00 \\ 
        & scancode$^{\dag}$ & 100.00 & 33.06 & 2.94 & 33.06 & 33.06 & 31.54 & 0.01 & 100.00 \\ 
        & cdxgen$^{\dag}$ & 100.00 & 90.87 & 0.03 & 90.87 & 18.76 & 90.87 & 0.00 & 100.00 \\ 
        \midrule
        \multirow{6}{*}{\rotatebox{90}{\textbf{Python}}} & syft & 100.00 & 100.00 & 1.62 & 42.64 & 0.95 & 42.78 & 0.00 & 100.00 \\ 
        & gh-sbom & 100.00 & 100.00 & 0.00 & 67.87 & 23.15 & 100.00 & 0.00 & 100.00 \\ 
        & ort & 100.00 & 100.00 & 0.00 & 21.42 & 22.91 & 67.49 & 82.13 & 24.40 \\ 
        & sbom-tool & 100.00 & 100.00 & 62.06 & 99.96 & 30.05 & 99.96 & 0.00 & 100.00 \\ 
        & scancod$^{\dag}$ & 100.00 & 50.86 & 39.24 & 50.86 & 50.86 & 31.40 & 0.38 & 100.00 \\ 
        & cdxgen$^{\dag}$ & 100.00 & 74.60 & 23.18 & 74.60 & 27.54 & 74.60 & 0.00 & 100.00 \\ 
        \bottomrule
    \end{tabular}
    }
    
    \raggedright\footnotesize
    ~$^{\dag}$ Tools that report results in CycloneDX standard, the unmarked tools are reported in SPDX standards.
\end{table}

\subsubsection{Results analysis} We analyze both SBOM tools and data fields for the compliance results.  

Data fields that can be directly extracted or derived from the software exhibit higher compliance rates. 
All mandatory data fields from SBOM standards, such as \textit{bom format} and \textit{spec version}, are automatically generated as metadata about the SBOM document itself. This creates a dangerous illusion of completeness: tools can be ``fully compliant'' with standards while providing incomplete or inaccurate software metadata. However, this can be a dangerous illusion of completeness that SBOM tools do not actually comply with the policy requirements.

Beyond this SBOM-intrinsic information, tools retrieve data to populate other SBOM fields from two primary sources: first, the software metadata files, such as ``requirements.txt'' in Python, or analysis results of Software Composition Analysis (SCA) technologies~\cite{imtiaz2021comparative}.  
Second, the external or online information, such as official websites or online package managers.
Fields like \textit{package name} and \textit{package version} are typically retrieved from metadata files, which are well-supported by all SBOM tools, leading to high compliance. Similarly, \textit{purl} can be derived from \textit{package name} and \textit{package version}, resulting in good compliance across tools.  
In contrast, fields like \textit{license}, \textit{supplier}, and \textit{copyright} often require SBOM tools to retrieve information from external resources.
For instance, gh-sbom and cdxgen implement online lookup mechanisms to fetch such data. Tools without these capabilities, such as syft, exhibit lower compliance for these fields. This indicates that the limited information gained directly from the software is not enough to fulfill the SBOM. SBOM tools must leverage online resources to formulate fully policy-compliant SBOMs.

Compared to prior studies such as~\cite{10.1145/3605098.3635927,cofano2024sbom,xiao2025jbomaudit}, we further report poor compliance of data fields within packages (\textit{e.g.}, \textit{license}, \textit{supplier}) and identifies discrepant gaps across languages, thereby highlighting the potential for ecosystem-specific biases that may be overlooked in analyses confined to single-language environments. We further found that the tools provided incomplete support for governmental policy requirements. Our new findings suggest the need for acquiring external information resources to gain more complete policy compliance with SBOM.


\begin{takeaway}
\textbf{Takeaways from RQ1}: 
Achieving 100\% compliance on a standard's mandatory fields creates a dangerous \textit{illusion of completeness}. This focus on minimal schema adherence masks critical gaps in policy-mandated data, \textit{e.g.}, \textit{NTIA}$^+$, exposing organizations to significant, unforeseen compliance risks. The lesson is that practical utility requires data completeness on the needed data fields, not just superficial validation on the standard mandatory data fields.
\end{takeaway}

\subsection{RQ2. Consistency}
\label{sec:consistency-results}
\subsubsection{Evaluation settings} 
The consistency evaluation builds on the compliance results. 
To ensure valuable evaluations, we first filter target data fields.
Fields in the mandatory group are excluded as they identify the SBOM itself and do not include software-specific information.
Similarly, \textit{creator} (of the SBOM, \textit{i.e.}, the tool) and \textit{timestamp} (of the SBOM) from the $NTIA^+$ are excluded as they are irrelevant for comparing among tools. The consistency evaluation focuses on direct information about software, thus excluding the inter-component analysis of \textit{package relationship}~consistency. 

\subsubsection{Consistency results} We employ \my to assess consistency between tool pairs within each standard.
We report the results of \textit{package name}, \textit{package version}, \textit{purl}, and \textit{license} in Table~\ref{tab:consistency}, the consistency score calculation of each data field is illustrated at Formula~(\ref{eqa:consistency-field}). 
The \textit{supplier} and \textit{copyright} fields show near-0\% consistency scores across tool pairs.
This occurs because only sbom-tool and ort adequately support them, while other tools provide nearly no support (as shown in compliance results Table~\ref{tab:compliance-data}). Therefore, we do not report these fields in Table~\ref{tab:consistency}. This complete inconsistency shows the vital importance of the SBOM compliance as we discussed in \S \ref{sec:compliance-results}. Without compliance, none of the downstream applications can be accomplished.

\begin{table*}[b]\small 
    \centering
    \caption{Consistency evaluation results. 
    The pkg., ver., and lic. represents the package, version, and license. The \textbf{bold} values are the best consistency scores of the tool pairs for each data field column within each standard. The package consistency is calculated by Formula~(\ref{eqa:consistency-pkg}), and the consistency of the other data fields is calculated with Formula~(\ref{eqa:consistency-field}). Reporting results in percentages.
    }
    \label{tab:consistency}
    \resizebox{\linewidth}{!}{
    \begin{tabular}{rrrrrrrrrrrrrr} 
        \toprule
        \multicolumn{2}{c}{} & \multicolumn{4}{c}{\textbf{C/C++}} & \multicolumn{4}{c}{\textbf{Java}} & \multicolumn{4}{c}{\textbf{Python}} \\
        \cmidrule[0.5pt](lr){3-6} \cmidrule[0.5pt](lr){7-10} \cmidrule[0.5pt](lr){11-14}
        & \textbf{Tool} & \textbf{pkg.} & \textbf{ver.} & \textbf{purl} & \textbf{lic.}  & \textbf{pkg.} & \textbf{ver.} & \textbf{purl} & \textbf{lic.} & \textbf{pkg.} & \textbf{ver.} & \textbf{purl} & \textbf{lic.}\\
        \midrule
        \multirow{10}{*}{\rotatebox{90}{\textbf{CycloneDX}}} & syft$\leftrightarrow$gh-sbom & 9.15 & 19.13 & 19.41 & 0.09 & 45.29 & 60.84 & 72.55 & 0.33 & 23.87 & 43.92 & 44.15 & 0.13 \\ 
        & syft$\leftrightarrow$ort & 1.21 & 3.08 & 2.87 & 0.00 & 5.94 & 14.51 & 14.49 & 0.12 & 3.05 & 6.46 & 6.14 & 0.00 \\ 
        & syft$\leftrightarrow$scancode & 0.41 & 4.89 & 5.44 & 0.95 & 3.02 & 16.67 & 16.91 & 0.60 & 0.42 & 2.50 & 2.65 & 0.77 \\ 
        & syft$\leftrightarrow$cdxgen & 12.05 & 28.48 & 23.31 & 0.11 & \textbf{63.90} & 85.73 & 71.99 & 0.32 & 17.24 & 34.97 & 34.54 & 0.19 \\ 
        & gh-sbom$\leftrightarrow$ort & 3.54 & 9.28 & 8.89 & 1.71 & 18.32 & 56.64 & 57.49 & \textbf{30.77} & 9.69 & 19.19 & 20.54 & 3.90 \\ 
        & gh-sbom$\leftrightarrow$scancode & 0.18 & 3.07 & 3.00 & 0.35 & 2.42 & 28.65 & 29.89 & 0.57 & 0.17 & 1.23 & 2.24 & 0.34 \\ 
        & gh-sbom$\leftrightarrow$cdxgen & \textbf{45.79} & \textbf{72.60} & \textbf{82.12} & \textbf{1.74} & 34.74 & \textbf{86.54} & \textbf{80.67} & 29.97 & \textbf{32.83} & \textbf{70.10} & \textbf{72.85} & \textbf{9.22} \\ 
        & ort$\leftrightarrow$scancode & 0.09 & 0.35 & 0.32 & 0.14 & 0.03 & 0.65 & 0.72 & 0.16 & 0.01 & 0.12 & 0.12 & 0.00 \\ 
        & ort$\leftrightarrow$cdxgen & 2.98 & 5.79 & 5.52 & 0.26 & 11.39 & 14.34 & 12.02 & 6.11 & 14.88 & 24.57 & 24.30 & 2.93 \\ 
        & scancode$\leftrightarrow$cdxgen & 0.33 & 4.61 & 4.44 & 0.66 & 1.66 & 15.62 & 13.35 & 3.73 & 0.29 & 2.95 & 3.30 & 0.50 \\ 
        \midrule
        \multirow{6}{*}{\rotatebox{90}{\textbf{SPDX}}} & syft$\leftrightarrow$gh-sbom & \textbf{32.41} & \textbf{41.98} & \textbf{46.08} & 0.00 & \textbf{7.98} & \textbf{16.75} & \textbf{18.36} & 0.00 & \textbf{28.23} & \textbf{45.84} & \textbf{45.68} & 0.00 \\ 
        & syft$\leftrightarrow$ort & 0.44 & 2.86 & 2.85 & 0.00 & 2.42 & 14.34 & 14.21 & 0.01 & 1.13 & 6.16 & 6.20 & 0.00 \\ 
        & syft$\leftrightarrow$sbom-tool & 7.27 & 12.85 & 12.80 & 0.00 & 3.36 & 5.65 & 5.62 & 0.00 & 9.58 & 22.01 & 22.38 & 0.00 \\ 
        & gh-sbom$\leftrightarrow$ort & 0.39 & 2.91 & 3.11 & 0.00 & 0.09 & 0.57 & 0.57 & 0.00 & 0.88 & 5.19 & 5.30 & 0.00 \\ 
        & gh-sbom$\leftrightarrow$sbom-tool & 6.51 & 12.77 & 13.23 & \textbf{8.51} & 3.07 & 5.42 & 5.58 & \textbf{4.33} & 8.55 & 21.75 & 22.26 & \textbf{18.24} \\ 
        & ort$\leftrightarrow$sbom-tool & 2.70 & 5.42 & 5.44 & 0.00 & 0.64 & 1.01 & 1.01 & 0.00 & 10.56 & 15.18 & 17.02 & 0.00 \\
        \midrule
        & \textbf{Average$^{\dag}$} & \textbf{7.84} & \textbf{14.38} & \textbf{14.93} & \textbf{0.91} & \textbf{12.77} & \textbf{26.50} & \textbf{25.96} & \textbf{4.81} & \textbf{10.09} & \textbf{20.13} & \textbf{20.60} & \textbf{2.26} \\ 
        \bottomrule
    \end{tabular}
    }
    
    \raggedright\footnotesize
    ~$^{\dag}$ Average consistency score of a data field across all tool pairs for each language.
    \vspace{-1em}
\end{table*}

\textbf{Overall results:} \textit{SBOM tools show poor overall consistency.}
Table \ref{tab:consistency} shows the variations in consistency levels among SBOM tools.
For instance, gh-sbom and cdxgen in CycloneDX achieve 10 of the highest consistency scores among tool pairs.
In contrast, the gh-sbom and sbom-tool pair in SPDX demonstrates no more than 25\% consistency across all data fields.
Both pairs include gh-sbom; their consistency differs, indicating a gap in the interchangeability of SBOMs.

Overall, the consistency of package detection is notably low, with average scores at 7.84\%, 12.77\%, and 10.09\% across languages, as detailed in Table~\ref{tab:consistency}.
For package-related fields, higher consistency is observed in those that can be directly retrieved from metadata files or derived, such as \textit{package version} and \textit{purl}, while \textit{license} exhibits poor consistency. These poor consistencies across SBOM tools break the interchangeability of SBOMs and prevent SBOMs from being substituted across organizations. This will result in unreliable SBOMs in scenarios like exchanging software vulnerabilities or managing software dependencies.

\begin{figure}[t]
  \vspace{-1em}
  \centering
  \includegraphics[width=0.7\linewidth]{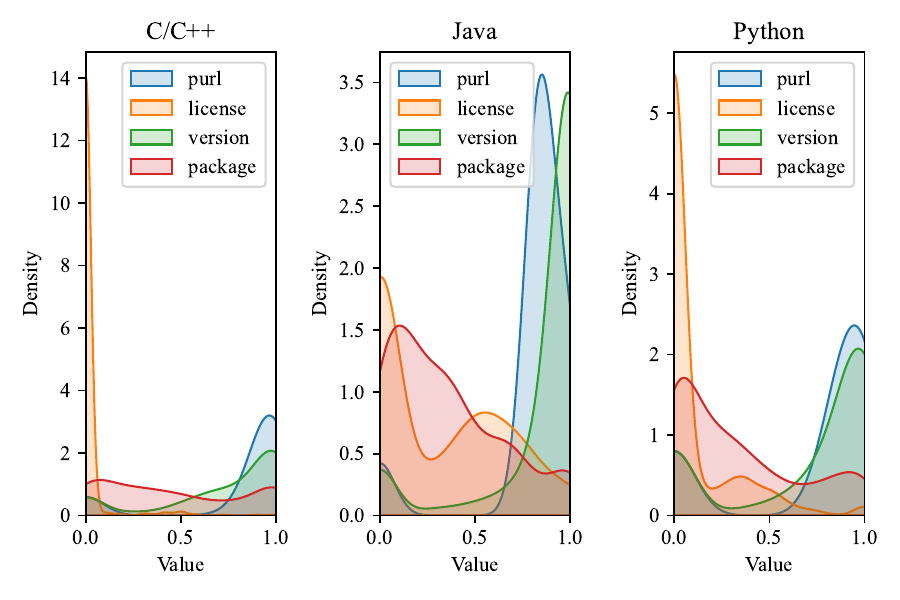}
  \caption{The Kernel Density Estimate (KDE) distribution of consistency results for all paired repositories of the gh-sbom$\leftrightarrow$cdxgen tool pair in the CycloneDX standard. The distribution position with high density means high portion of repositories show the consistency score at that position. Best viewed in color.}
  \Description{The figure contains three subplots corresponding to C/C++, Java, and Python repositories, respectively. Each subplot shows KDE curves for four consistency metrics: purl (blue), license (orange), version (green), and package (red). The x-axis represents the consistency score value ranging from 0 to 1, while the y-axis indicates the density, reflecting the proportion of repositories exhibiting that score. Peaks in the curves indicate a higher frequency of repositories with that specific consistency score.}
  \label{fig:detail_consistency}
\end{figure}

\textbf{Across languages:} 
Based on the average results in the last row of Table \ref{tab:consistency}, tool pairs exhibit higher consistency in Java and Python compared to C/C++.
To understand this trend in detail, we analyze the Kernel Density Estimate (KDE) distribution of consistency scores for various data fields from gh-sbom and cdxgen using CycloneDX, as shown in Figure~\ref{fig:detail_consistency}.
The distribution shows a relatively higher density of ``high consistency'' scores for Java and Python, especially in the package detection.
The results align with the package manager support for Java, Python, and C/C++, which is used by SBOM tools to extract relevant data fields. 
This result shows a reliance on the metadata of package managers by SBOM tools for detecting dependent packages and suggests that SBOM tools may not produce complete SBOMs in environments that have poor package manager support.

\textbf{Across SBOM standards:} Tools supporting both SPDX and CycloneDX (gh-sbom, ort, syft) can generate self-inconsistent outputs for the same project when switching between standards, undermining interchangeability. 
Take the ort and syft pair analyzing Java repositories as an example, none of the data fields gain the same consistency scores, similar for all the other tool pairs. This inconsistency can result in problems when migrating SBOMs from different SBOM standards, which should be overcome in the design of SBOM tools.
Based on our analysis, this can occur from: 

(1)~Structural differences between standards, \textit{e.g.}, SPDX includes the primary software as a package within the same section as its dependencies, unlike CycloneDX, which uses an independent ``metadata'' section.
This nearly halves ort and syft's Java package consistency in SPDX as~Formula~(\ref{eqa:consistency-pkg}).

(2)~The use of different data acquisition APIs for different standards by tools like gh-sbom (\textit{e.g.}, GitHub Dependency Graph for SPDX vs. GraphQL/ClearlyDefined for CycloneDX), also leads to divergent results. For instance, gh-sbom's C/C++ \textit{package} consistency with syft drops from 32.41\% (SPDX) to 9.15\% (CycloneDX). Meanwhile, the consistency score of Java \textit{license} field in gh-sbom and ort changes from 0.00\% (SPDX) to 30.77\% (CycloneDX).

\subsubsection{Inconsistency issues} 
We analyze individual instances observed during the evaluation and identify several critical types of inconsistency issues, which are detailed below.
\label{sec:inconsis-taxo}

\textbf{Inconsistent organization of information:} We observe that SBOM tools may adopt different approaches to organizing the same data fields.
For instance, scancode stores the \textit{license} field as a plain string, such as ``Apache-2.0 AND MIT'', while gh-sbom organizes it as a logical dictionary field, like: ``\{`licenses':\{`license':\{`id': `Apache-2.0', `text': `xxxx'\}\}\}''. 
    Such organizational inconsistencies can hinder interchangeability in the scenario of open source license management and lead to inaccuracies in the license conflict detection scenario.
\my implements specific extraction methods for these varied license representations into a consistent, comparable format, \textit{i.e.}, by converting expressions into a canonical list of individual license identifiers as detailed in \S \ref{sec:evaluation-cons}.

\textbf{Inconsistent description methods:} 
SBOM tools may employ different methods to describe the same information.
For example, some tools add a package manager prefix, such as recording ``pytorch-lighting'' as ``pip:pytorch-lighting'' in SBOMs generated by gh-sbom. 
Other identifiers like ``maven'', ``docker'', and ``actions'' are also frequently observed. 
While these prefixes can aid in identifying the information source, they may also introduce inconsistencies in exchange for the same packages across SBOMs.
\my addresses this by standardizing the names, \textit{i.e.}, removing prefixes associated with a set of widely used package managers through regular expression matching.

\textbf{Inconsistent content:} \my has addressed the aforementioned discrepancies to ensure effectiveness, yet the low consistency of SBOM tools in content remains evident, as shown in Table \ref{tab:consistency}.
To further investigate fields within packages, \my is also employed to examine other SBOM sections, such as ``file'' information in SPDX SBOMs.
This information is not reported in Table \ref{tab:consistency} because CycloneDX does not support the ``files'' section.

We identify some surprising anomalies: SBOM tools provide inconsistent checksums for the same file within software calculated using the same hash algorithm, as shown in Table \ref{tab:checksums}. 
We further inspect the SBOM and found that ort reports three \textit{checksums} for the same filename in both cases. 
However, corresponding checksums, \textit{i.e.}, checksums that should match those from ort for the same file content, are present in sbom-tool's SBOM for the first case but are absent for the second.
This situation reveals at least two issues: 
(1)~the absence of a unified file naming strategy, 
and (2)~the lack of clear instructions for reproducing and validating \textit{checksums}. 
These issues highlight how varying interpretations of the standard guidelines by tool developers can cause such inconsistencies. This phenomenon is also quite similar to the packages with identical package names and versions that we addressed with the triple-factor best-matching method in \S \ref{sec:evaluation-cons}, which shows the practical bad effect on the ambiguous mispaired packages.

\begin{table}[b]\small
    \centering
    \caption{Checksums inconsistency instances for files within ``redcarpet'' and ``rugged'' with ``SHA-1'' algorithm.}
    \label{tab:checksums}
    \begin{tabular}{rrrr}
        \toprule
        \textbf{Tool} & \textbf{Repository} & \textbf{Filename} & \textbf{Checksums} \\
        \midrule
        sbom-tool & redcarpet & ./COPYING & 82f5b22dbc4dbd63320f7442109268140d72168f \\
        ort & redcarpet & COPYING & 0301cb36cb4c34ab1a92a18949843cebe02cec71 \\
        sbom-tool & rugged & ./Rakefile & bb1089ca7ec532481bb5744326b0c3443feb1625 \\
        ort & rugged & Rakefile & d9c843fa4ec89f3f8207bd403d21f198e97e812b \\
        \bottomrule
    \end{tabular}
    \label{tab:inconsistency-checksums}
    \vspace{-1em}
\end{table}

\textbf{Longitudinal Analysis: the persistence and volatility of adherence gaps} A critical question arising from our baseline analysis is whether the observed adherence gaps are transient issues of a nascent ecosystem or persistent, systemic challenges. To address this, we conducted a one-year longitudinal follow-up study, repeating our entire experimental procedure with the latest stable tool versions as of October 2025\footnote{~The updated versions are: cdxgen@v11.8.0, gh-sbom@v0.1.1, ort@v69.0.0, syft@v1.33.0, sbom-tool@v4.1.1, and scancode@v32.4.1. The target SBOM standards (SPDX 2.3 and CycloneDX 1.5) remained the same. In summary, we gain 27,649 SBOMs at this stage, 91.61\% (25330) of which are successfully processed.}.
This two-stage approach allows us to measure the temporal stability of the gaps and quantify the impact of tool evolution itself.

Our longitudinal analysis yields two critical insights that both reinforce our baseline findings and expose a new dimension of the ecosystem's instability.

\textit{(1) Tool evolution is a primary driver of inconsistency:} We first investigated the consistency of each tool with its \textit{own} prior version to isolate the impact of evolution. The results, presented in Table~\ref{tab:consistency_same}, are striking. The average intra-tool consistency for \textit{package detection} is only 51.62\% for C/C++, 52.61\% for Java, and a mere 46.96\% for Python. This demonstrates that tool evolution is a major, and perhaps under-appreciated, driver of instability in the SBOM ecosystem. An organization that simply updates its tooling may find its new SBOMs are massively inconsistent with its historical data, breaking downstream security and compliance workflows. The effect is also highly volatile: while tools like gh-sbom (in CycloneDX) and sbom-tool show high self-consistency (often >90\%), ort exhibits extremely low self-consistency (<15\% in many cases), despite its major version update. This finding empirically validates our decision to treat different tool versions as functionally distinct entities in our consistency analysis.

\textit{(2) The systemic inter-tool gap persists and remains critically low:} Beyond intra-tool evolution, our 2025 follow-up confirms that the poor \textit{inter-tool} consistency observed in our baseline is a persistent systemic issue. The average inter-tool consistency for package detection in our follow-up evaluations was 6.77\% for C/C++, 8.70\% for Java, and 8.42\% for Python (detailed statistics are provided in Table~\ref{app-tab:consistency} within Appendix). These figures show no statistically significant improvement from our 2024 baseline (7.84\%, 12.77\%, and 10.09\% respectively). This finding strongly refutes any notion that the adherence gap is a temporary problem that is rapidly resolving itself. Instead, it indicates that the deep-rooted challenges identified in our baseline analysis, such as ambiguous standard interpretations and divergent tool implementations, remain the primary obstacles to achieving an interchangeable and reliable SBOM ecosystem.

\begin{table*}[t]
    \vspace{-1em}
    \centering
    \caption{Longitudinal consistency evaluation results on different versions of the same tool. The pkg., ver., and lic. represent for the package, version, and license. The package consistency is calculated by Formula~(\ref{eqa:consistency-pkg}), and the consistency of the other data fields is calculated with Formula~(\ref{eqa:consistency-field}). Reporting results in percentages.}
    \label{tab:consistency_same}
    \resizebox{\linewidth}{!}{
    \begin{tabular}{rrrrrrrrrrrrrr} 
        \toprule
        \multicolumn{2}{c}{} & \multicolumn{4}{c}{\textbf{C/C++}} & \multicolumn{4}{c}{\textbf{Java}} & \multicolumn{4}{c}{\textbf{Python}} \\
        \cmidrule[0.5pt](lr){3-6} \cmidrule[0.5pt](lr){7-10} \cmidrule[0.5pt](lr){11-14}
        & \textbf{Tool$^{\dag}$} & \textbf{pkg.} & \textbf{ver.} & \textbf{purl} & \textbf{lic.}  & \textbf{pkg.} & \textbf{ver.} & \textbf{purl} & \textbf{lic.} & \textbf{pkg.} & \textbf{ver.} & \textbf{purl} & \textbf{lic.}\\
        \midrule
        \multirow{5}{*}{\rotatebox{90}{\textbf{CycloneDX}}} & \textbf{cdxgen} & 58.19 & 64.77 & 65.21 & 2.02 & 80.71 & 89.48 & 89.69 & 0.68 & 41.26 & 59.01 & 63.01 & 14.64 \\ 
        & gh-sbom & 86.62 & 92.72 & 98.56 & 9.74 & 92.94 & 95.93 & 99.01 & 41.87 & 89.96 & 94.80 & 98.70 & 19.14 \\ 
        & \textbf{ort} & 3.00 & 3.01 & 3.09 & 3.00 & 4.17 & 4.14 & 4.21 & 5.49 & 9.36 & 9.88 & 10.61 & 14.06 \\ 
        & scancode & 18.83 & 17.71 & 23.49 & 23.72 & 32.51 & 31.80 & 32.97 & 32.55 & 48.13 & 30.53 & 49.31 & 48.80 \\ 
        & syft & 45.56 & 59.54 & 55.52 & 1.10 & 57.26 & 93.18 & 92.91 & 2.50 & 38.84 & 50.61 & 49.63 & 1.10 \\ 
        \midrule
        \multirow{4}{*}{\rotatebox{90}{\textbf{SPDX}}} & gh-sbom & 42.26 & 52.40 & 59.66 & 9.05 & 43.29 & 39.59 & 50.64 & 22.32 & 36.73 & 58.29 & 42.19 & 22.69 \\ 
        & \textbf{ort} & 80.96 & 94.06 & 2.56 & 1.60 & 14.67 & 19.15 & 3.79 & 4.20 & 60.55 & 64.42 & 9.15 & 10.58 \\ 
        & \textbf{sbom-tool} & 90.76 & 99.99 & 91.75 & 4.10 & 97.17 & 99.99 & 92.09 & 3.36 & 64.93 & 100.00 & 91.80 & 3.17 \\ 
        & syft & 45.56 & 1.10 & 59.54 & 55.52 & 57.26 & 2.50 & 93.18 & 92.91 & 38.84 & 1.10 & 50.61 & 49.63 \\ 
        \midrule
        & \textbf{Average$^{\ddag}$} & 51.62 & 60.42 & 50.69 & 6.17 & 52.61 & 62.94 & 62.03 & 12.94 & 46.96 & 57.57 & 51.58 & 15.04 \\ 
        \bottomrule
    \end{tabular}
    }
    
    \raggedright\footnotesize
    ~$^{\dag}$ The tools with \textbf{bold} have major version updates, \textit{e.g.}, from \textbf{[10]}.10.4 updates to \textbf{[11]}.8.0 of the cdxgen.\\
    ~$^{\ddag}$ Average consistency score of a data field across all tool pairs for each language.
    
\end{table*}

\subsubsection{Results analysis}

We analyze both SBOM standards and SBOM tools to identify the root causes of the inconsistencies.

\textbf{Unclear standard restrictions and validation:}
The flexibility in SBOM standards, intended for broad adoption, may lead to vary interpretations by developers. This results in inconsistencies, such as improper use of the \textit{package name} field, \textit{e.g.}, for package manager details instead of a dedicated source field, or varying \textit{checksums} for the same file due to inadequate guidance on file identification and reproduce steps, reflecting issues as inconsistent contents or description methods.

\textbf{Ambiguous SBOM scope definition:} Lack of dedicated scope fields in standards allows tools to generate SBOMs with varying scopes, causing content inconsistencies for the same software.
\begin{itemize}
    \item \textit{Language-level scope:} A tool's capability to report packages from all programming languages used in a project dictates its effective SBOM scope, leading to variance and comparison challenges when multi-language reporting differs.
    \item \textit{Package-level scope:} Standards' unclear directives on including transitive dependencies cause tools to differ. Some may list only direct dependencies, while others, \textit{e.g.}, those with online access, include transitive ones, resulting in disparate dependency depths and inconsistent~scopes.
\end{itemize}

Our work illustrates poor package-level consistency, similar to \cite{yu2024correctness}, but advances further by detailing field-level inconsistencies within software packages, \textit{e.g.}, average \textit{version} and \textit{purl} consistencies under 26.50\% across languages.
Moreover, this study is, to our knowledge, the first to systematically uncover inconsistencies in SBOMs generated by the same tools for different~standards.


\begin{takeaway}
\noindent\textbf{Takeaways from RQ2}: 
The extremely low inter-tool consistency, with package detection agreement as low as 7.84\% to 12.77\% on average, means that current SBOMs produced by SBOM tools are not interchangeable. The inconsistency between evolved SBOM tool versions even exacerbated the problem. This failure breaks the chain of trust in any cross-organizational workflow and undermines the core purpose of a standardized format. The key insight is that this is not just a tooling failure but a symptom of ambiguous standards that require clarification.
\end{takeaway}

\subsection{RQ3. Accuracy}
\label{sec:accuracy-results}

\subsubsection{Evaluation settings} Following the broad multi-language evaluations of compliance and consistency, this section evaluates the accuracy of SBOM tools in capturing essential software component information, using the Python ground truth dataset $D_{gt}$. This focused approach allows for a deep dive into the nuances of accuracy challenges within a well-understood ecosystem.

Note that the $D_{gt}$ does not contain the field of \textit{purl}. 
Therefore, the triple-factor best-matching algorithm used in the consistency evaluation is reduced to matching the best pair of \textit{package name} and \textit{package version}, requiring an exact match for the package name ($\tau = 1.0$) to establish a correspondence. 
Once a package name is matched, the version string is evaluated by the Formula~(\ref{eqa:consistency-field}). 
All other data fields are evaluated using the same methodology as in the consistency study.

A critical nuance in our accuracy assessment concerns missing version information. As shown in Table~\ref{tab:bench}, 74.4\% of the dependent packages in $D_{gt}$ do not declare an explicit version. 
In the context of our evaluation, when the ground truth for a package's version is absent, a tool is considered accurate on the \textit{version} field only if its generated SBOM also reports a version as an empty string or equivalent null value. 
This strict criterion is applied because, lacking a ground truth version, the presence of any non-empty version string in the SBOM cannot be verified as correct and may represent an inferred or default value. 
This design ensures that the evaluation reflects the tool's fidelity to the actual (incomplete) state of the software, rather than rewarding speculation.

\subsubsection{Accuracy results}
\textbf{Overall results:} SBOM tools exhibit moderate accuracy in package detection and generally perform poorly when identifying data fields within packages.
Considering the accurate construction process of $D_{gt}$, it should be expected that SBOM tools would maintain high ``recall'' scores in package detection, \textit{i.e.}, should not miss the evident dependencies.
However, as shown in Table \ref{tab:acc}, only gh-sbom and cdxgen demonstrate relatively strong performance in detecting packages, with recall rates of 58.50\% and 42.96\%, respectively.
The gh-sbom also achieves 66.6\% accuracy for the \textit{package version} field. Other tools generally exhibit significantly lower performance. 
Moreover, fields within packages such as \textit{supplier} and \textit{license} exhibit poor accuracy across SBOM tools.
This overall poor accuracy of SBOM tools will prevent the applications that rely on SBOMs, such as vulnerable dependency detection or open source license management.

\begin{table}[t]\small
    \centering
    \vspace{-1em}
    \caption{Accuracy evaluation results. ``Count'' means the evaluated valid SBOMs. ``--'' means the evaluated SBOM tool finds no packages in $D_{gt}$.}
    \label{tab:acc}
    \centering
    \begin{minipage}{0.76\linewidth}
    \centering
    \begin{tabular}{rrrrrrr}
        \toprule
        \textbf{Tool} & \textbf{Count} & \textbf{Precision$^{\dag}$} & \textbf{Recall$^{\dag}$} & \textbf{supplier} & \textbf{license} & \textbf{version} \\ 
        \midrule
        syft & 100 & 17.69\% & 12.39\% & 0.00\% & 0.00\% & 18.00\% \\ 
        gh-sbom & 36 & 73.84\% & 58.50\% & 0.00\% & 8.72\% & 64.30\% \\ 
        ort & 84 & 12.01\% & 16.90\% & 0.00\% & 13.95\% & 5.95\% \\ 
        scancode & 100 & -- & -- & -- & -- & -- \\ 
        cdxgen & 98 & 29.78\% & 42.96\% & 18.15\% & 14.76\% & 15.31\% \\ 
        sbom-tool$^{\ddag}$ & 100 & 14.50\% & 24.38\% & 9.69\% & 18.49\% & 11.07\% \\ 
        \bottomrule
    \end{tabular}
    
    \raggedright\footnotesize
    $^{\dag}$ The precision and recall are calculated on package detection as in \S \ref{sec:acc-design}.\\
    $^{\ddag}$ The sbom-tool is in SPDX, the other tools are in CycloneDX.
    \end{minipage}
\end{table}

\textbf{Relation with consistency:}
Tools with higher consistency generally exhibit higher accuracy. 
For instance, cdxgen and gh-sbom achieve 32.83\% consistency in package detection for Python (Table \ref{tab:consistency}), and they also demonstrate strong precision and recall, as discussed above.
Similarly, data fields with low consistency also exhibit low accuracy, \textit{e.g.}, the \textit{license} field. 
In addition to these overall trends, the inconsistency issues described in \S \ref{sec:inconsis-taxo} also contribute to inaccuracies in SBOMs. These findings underscore that consistency is not merely a desirable quality but a practical indicator of an SBOM tool's reliability, serving as a key criterion for tool selection and evaluation.

Notably, scancode achieves 0\% accuracy on $D_{gt}$ as the {``--''} in Table~\ref{tab:acc}, which is consistent with its low inter-tool consistency for Python (Table~\ref{tab:consistency}).
This is observed even though \my adopts all the officially recommended parameters (\textit{i.e.}, ``cpeui'') for its generation.
Examination of all the scancode SBOMs for $D_{gt}$ reveals that 50 of them list only the repository itself as a package, while the remaining 50 SBOMs contain only the information of the scancode itself.
This indicates a potential limitation of the tool, resulting in no matches against $D_{gt}$ entries.

\subsubsection{Results analysis}

\begin{figure}[t]
  \vspace{-1em}
  \centering
  \includegraphics[width=0.7\linewidth]{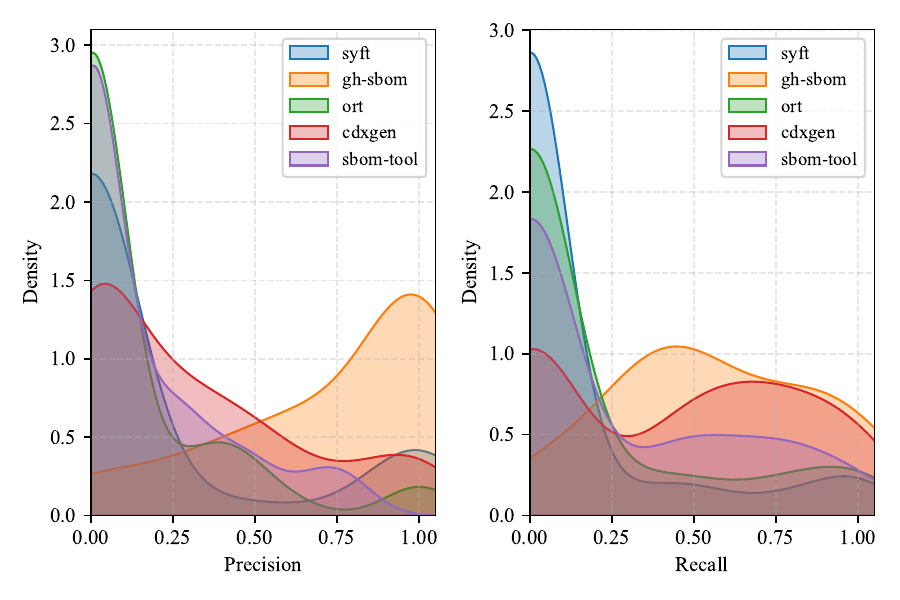}
  \caption{The KDE distribution of precision and recall results for package detection across SBOM tools. The sbom-tool uses the SPDX format, while the other tools use CycloneDX. This result illustrates the relatively high proportion of mismatched packages in the SBOMs generated by SBOM tools. Best viewed in color.}
  \Description{The left subplot shows the density distribution of precision scores, reflecting the accuracy of package identification by different tools. The right subplot presents the density of recall scores, indicating the tools' ability to detect all packages. Different colors correspond to various SBOM tools, highlighting their comparative performance.}
  \label{fig:detail_acc}
\end{figure}

\textbf{Language and package scope:} 
Consistent with the analysis in the consistency results, the unclear language scope of SBOMs can also lead to inaccuracies in package detection.
As illustrated by the 0\% accuracy of package detection by scancode, the lack of a well-defined SBOM scope can result in significant utilization challenges for software supply chain management.
Beyond the capabilities of SBOM tools, the unclear scope of transitive dependencies can also contribute to low precision.
The analysis of the distribution of package detection results across tools (Figure~\ref{fig:detail_acc}) reveals a higher proportion of ``low scores'' in precision compared to recall. 
This aligns with the inclusion of evident dependencies in $D_{gt}$, which may cause false positives.

\textbf{Substandard information within repositories:} 
SBOM tools rely on metadata files within repositories to detect packages. 
However, non-standard formats can occur, such as ``install.txt'' instead of ``requirements.txt'' in Python, which can hinder the tools' ability to extract dependencies accurately. 
Additionally, software dependencies may have import names that differ from their package names. 
For example, the package ``opencv-python'' is imported as ``cv2''. 
If not properly checked or handled by the tools, they may retrieve incorrect data.

\textbf{Missing and unstructured information:}
Software repositories often lack explicit dependency specifications, which can lead to incomplete SBOM generation. 
As observed during the construction of $D_{gt}$ in \S \ref{sec:gt}, 74.4\% of detected dependent packages did not declare explicit version information. 
To address such missing metadata, some SBOM tools (\textit{e.g.}, cdxgen) interact directly with language-specific package managers or even attempt to install software dynamically to infer dependency details.
However, this reactive, online interaction introduces two further issues:
(1) it may create a false sense of dependency security, as the SBOM tool generates versioned entries that were not actually pinned by developers, and (2) online platforms themselves may contain incomplete or inaccurate metadata, thereby compromising the reliability of the retrieved information.
Additionally, even when tool-fetched data is available, it can be inconsistently structured. 
For example, querying the PyPI API for the ``elasticsearch'' package may return ``none'' for \textit{license}, even though a license is clearly stated on the project's web page.

Compared to \cite{cofano2024sbom,10.1145/3605098.3635927,yu2024correctness,xiao2025jbomaudit}, our granular field-level analysis reveals the poor accuracy of SBOM tools in supplier and license extraction, \textit{i.e.}, all evaluated tools perform under 20\% accuracy for these fields.
Our real-world ground truth dataset $D_{gt}$ further exposes the limitations of tools in handling non-standard repositories; this is evidenced by a dependency detection recall rate of under 58.50\% from the evaluated tools. These evaluation results further underscore that the SBOMs produced by the SBOM tools are still not reliable for downstream applications.


\begin{takeaway}
\textbf{Takeaways from RQ3}: 
Poor accuracy, with rates below 20\% for critical fields like licenses, renders current SBOMs unreliable for automated security and compliance decisions. This is a ``garbage in, garbage out'' problem; the tools' failures are often a direct result of inconsistent and incomplete metadata within the scanned software projects themselves. Furthermore, the missing information within a single source can degrade the quality of the SBOM. Therefore, improving the accuracy of SBOM tools is fundamentally tied to improving repository quality and adding corroborative information sources.
\end{takeaway}

\subsection{Empirical evaluations on \my}

\label{sec:sap-empirical}

The effectiveness of the \my itself is of vital importance for the reliability of our evaluation results and corresponding findings. Thus, we empirically assess the three modules in the evaluation pipeline of \my to figure it out, and we report the empirical evaluation results in this section.

\subsubsection{Evaluations on the SBOM generation module}

As detailed in Section~\ref{sbom-generate}, the SBOM generation pipeline successfully produces 27,795 SBOM files, demonstrating its capability to process a large volume of repositories. Specifically, the sbom-tool and syft successfully generate SBOMs for every repository in $D_{repo}$, which demonstrates the effectiveness of \my's orchestration.

\subsubsection{Evaluations on the extraction module}

Next, we focus on evaluating the extraction module of \my (\S \ref{sec:extract_module}), which produces the extracted and organized JSON for the subsequent evaluations.

\begin{table}[t]\small
    \vspace{-1em}
    \centering
    \caption{Empirical evaluation results on \my's extraction module.}
    \label{tab:empirical}
    \begin{minipage}{0.64\linewidth}
    \centering
    \begin{tabular}{rrrrr}
        \toprule
        \textbf{Tool} & \textbf{C/C++} & \textbf{Java} & \textbf{Python} & \textbf{Total} \\
        \midrule
        \textbf{cdxgen} & 100.00\% & 100.00\% & 100.00\% & \textbf{100\%}(3243/3243) \\ 
        \textbf{sbom-tool$^{\dag}$} & 99.90\% & 100.00\% & 99.92\% & \textbf{99.94\%}(3285/3287) \\ 
        \textbf{scancode} & 100.00\% & 100.00\% & 100.00\% & \textbf{100\%}(3275/3275) \\ 
        \textbf{gh-sbom$^{\ddag}$} & 76.84\% & 66.15\% & 81.42\% & \textbf{74.64\%}(4713/6314) \\ 
        \textbf{ort$^{\ddag}$} & 100.00\% & 100.00\% & 100.00\% & \textbf{100.00\%}(5102/5102) \\ 
        \textbf{syft$^{\ddag}$} & 100.00\% & 99.73\% & 100.00\% & \textbf{99.91\%}(6568/6574) \\ 
        \bottomrule
    \end{tabular}
    
    \raggedright\footnotesize
    $^{\dag}$ Tool in SPDX, while the unmarked tools are in CycloneDX.\\
    $^{\ddag}$ Tools supporting both CycloneDX and SPDX.
    \end{minipage}
\end{table}

We evaluate the extraction module on all generated SBOMs.
Table~\ref{tab:empirical} shows that the module successfully extracts 100\% SBOMs into the defined structure from cdxgen, ort, and scancode, and over 99.9\% from sbom-tool and syft.
We inspect the eight failures: six from syft and one from sbom-tool are attributed to empty files produced by these tools, likely due to their internal generation errors.
The remaining one sbom-tool failure stems from an incomplete SBOM that triggered a \texttt{JSONDecodeError}.
Notably, \my achieves a 74.64\% success rate for gh-sbom. We found that all 1,601 failures correspond to empty files, caused by repositories disabling GitHub's ``Dependency Graph'' feature, which is required for gh-sbom generation. This also reveals the limitation of locally analyzing software by gh-sbom.

In summary, \my's extraction module successfully processes 26,186 SBOMs (94.2\% of the total 27,795), with failures primarily due to empty input files, confirming its reliability. The evaluation module leverages the extraction module's high-quality output, enabling robust subsequent analyses.

\subsubsection{Evaluations on the evaluation module} The foundation of the evaluation in the evaluation module of \my is the pairing of packages. As discussed in \S \ref{sec:evaluation_module}, we design the triple-factor best-matching algorithm for pairing packages in \my. This algorithm computes a matching score ranging from 0 to 3.0 (name: 0-1, version: 0-1, purl: 0-1), with the basic constraint that the package name must achieve a perfect match (score=1.0) before version and purl comparisons are considered valid. The threshold parameter $\tau$ determines the minimum acceptable matching score for establishing package correspondence, directly influencing the calculated consistency metrics.

Based on the definition of the triple-factor best-matching algorithm, the possible values of $\tau$ range from 1.0 to 3.0, corresponding with ``accept all matches with valid name pairing'' to ``perfect matching across all three factors''. Thus, we select five representative threshold values ($\tau\in\{1.0,1.5,2.0,2.5,3.0\}$) to systematically explore this parameter. This selection spans the full spectrum from the most permissive to the most restrictive matching criteria, with increments of 0.5 chosen based on pilot studies showing these values capture meaningful transitions in the trade-off.

\begin{table*}[t]\small
    \vspace{-1em}
    \centering
    \caption{Empirical evaluation results on the evaluation module of \my. The $\tau$ is the threshold for the triple-factor best-matching algorithm as defined in Formula~(\ref{eqa:match-2}). The numbers in the table are the package consistency score of SBOM tool pairs as calculated by Formula~(\ref{eqa:consistency-pkg}).
    All the results are reported in percentages.
    }
    \label{tab:tau-empirical}
    \resizebox{\linewidth}{!}{
    \begin{tabular}{rrrrrrrrrrrrrrrrr} 
        \toprule
        & & \multicolumn{5}{c}{\textbf{C/C++}} & \multicolumn{5}{c}{\textbf{Java}} & \multicolumn{5}{c}{\textbf{Python}} \\
        \cmidrule[0.5pt](lr){3-7} \cmidrule[0.5pt](lr){8-12} \cmidrule[0.5pt](lr){13-17}
        & \textbf{Tool pair} & \textbf{$\tau=1$} & \textbf{$\tau=1.5$} & \textbf{$\tau=2$} & \textbf{$\tau=2.5$}  & \textbf{$\tau=3$}& \textbf{$\tau=1$} & \textbf{$\tau=1.5$} & \textbf{$\tau=2$} & \textbf{$\tau=2.5$}  & \textbf{$\tau=3$} & \textbf{$\tau=1$} & \textbf{$\tau=1.5$} & \textbf{$\tau=2$} & \textbf{$\tau=2.5$}  & \textbf{$\tau=3$}\\
        \midrule
        \multirow{10}{*}{\rotatebox{90}{\textbf{CycloneDX}}} & syft$\leftrightarrow$gh-sbom & 9.24 & 9.24 & 9.15 & 9.01 & 6.58 & 48.12 & 48.12 & 45.29 & 37.51 & 33.71 & 23.95 & 23.95 & 23.87 & 23.71 & 22.85 \\
        & syft$\leftrightarrow$ort & 1.23 & 1.23 & 1.21 & 1.21 & 0.27 & 6.35 & 6.35 & 5.94 & 5.90 & 5.33 & 3.09 & 3.09 & 3.05 & 2.99 & 1.39 \\ 
        & syft$\leftrightarrow$scancode & 0.68 & 0.68 & 0.41 & 0.33 & 0.27 & 3.29 & 3.29 & 3.02 & 2.82 & 2.48 & 0.66 & 0.66 & 0.42 & 0.42 & 0.29 \\ 
        & syft$\leftrightarrow$cdxgen & 12.18 & 12.18 & 12.05 & 11.95 & 7.07 & 65.30 & 65.30 & 63.90 & 63.60 & 2.53 & 17.31 & 17.31 & 17.24 & 17.10 & 16.09 \\ 
        & gh-sbom$\leftrightarrow$ort & 5.12 & 5.12 & 3.54 & 3.43 & 0.36 & 22.42 & 22.42 & 18.32 & 17.62 & 14.87 & 19.76 & 19.76 & 9.69 & 7.12 & 2.08 \\ 
        & gh-sbom$\leftrightarrow$scancode & 0.22 & 0.22 & 0.18 & 0.16 & 0.06 & 3.42 & 3.42 & 2.42 & 2.27 & 1.55 & 0.24 & 0.24 & 0.17 & 0.12 & 0.00 \\ 
        & gh-sbom$\leftrightarrow$cdxgen & 48.35 & 48.35 & 45.79 & 38.64 & 31.71 & 40.49 & 40.49 & 34.74 & 32.02 & 13.73 & 50.45 & 50.45 & 32.83 & 30.38 & 22.96 \\ 
        & ort$\leftrightarrow$scancode & 0.12 & 0.12 & 0.09 & 0.09 & 0.02 & 0.03 & 0.03 & 0.03 & 0.02 & 0.00 & 0.04 & 0.04 & 0.01 & 0.01 & 0.01 \\ 
        & ort$\leftrightarrow$cdxgen & 2.98 & 2.98 & 2.98 & 2.98 & 1.94 & 11.39 & 11.39 & 11.39 & 11.39 & 0.17 & 14.93 & 14.93 & 14.88 & 14.76 & 11.81 \\ 
        & scancode$\leftrightarrow$cdxgen & 1.10 & 1.10 & 0.33 & 0.33 & 0.28 & 1.72 & 1.71 & 1.66 & 1.64 & 0.11 & 4.01 & 4.01 & 0.29 & 0.28 & 0.19 \\ 
        \midrule
        \multirow{6}{*}{\rotatebox{90}{\textbf{SPDX}}} & syft$\leftrightarrow$gh-sbom & 33.46 & 33.46 & 32.41 & 26.40 & 5.81 & 8.09 & 8.09 & 7.98 & 7.01 & 2.72 & 28.47 & 28.35 & 28.23 & 27.83 & 18.47 \\ 
        & syft$\leftrightarrow$ort & 0.63 & 0.56 & 0.44 & 0.43 & 0.41 & 2.71 & 2.63 & 2.42 & 2.36 & 2.19 & 1.16 & 1.16 & 1.13 & 1.07 & 1.01 \\ 
        & syft$\leftrightarrow$sbom-tool & 7.49 & 7.29 & 7.27 & 7.26 & 7.00 & 4.03 & 3.36 & 3.36 & 3.35 & 3.33 & 9.70 & 9.65 & 9.58 & 9.44 & 9.07 \\ 
        & gh-sbom$\leftrightarrow$ort & 0.81 & 0.81 & 0.39 & 0.38 & 0.34 & 0.16 & 0.16 & 0.09 & 0.09 & 0.08 & 4.17 & 0.88 & 0.88 & 0.86 & 0.81 \\ 
        & gh-sbom$\leftrightarrow$sbom-tool & 7.98 & 7.95 & 6.51 & 6.40 & 5.53 & 3.34 & 3.34 & 3.07 & 3.02 & 2.84 & 16.69 & 8.62 & 8.55 & 8.39 & 8.00 \\ 
        & ort$\leftrightarrow$sbom-tool & 76.21 & 2.83 & 2.70 & 2.69 & 2.64 & 10.27 & 0.64 & 0.64 & 0.64 & 0.58 & 54.78 & 11.96 & 10.56 & 9.94 & 5.11 \\
        \bottomrule
    \end{tabular}
}
\end{table*}

Table~\ref{tab:tau-empirical} shows the package consistency scores across $\tau$ values, which reveals several key~findings:

\begin{itemize}
    \item \textbf{Package pairing consistency scores decrease monotonically as $\tau$ increases:} As expected, higher thresholds reject more potential matches. However, the rate of decrease varies significantly across tool pairs. For the ort$\leftrightarrow$sbom-tool pair, its C/C++ consistency shows a sharp drop from 76.21\% to 2.83\% by the $\tau$ increase from 1.0 to 1.5, similarly for the Java ecosystem that drops from 10.27\% to 0.64\%. This is caused by the poor support of the ort on the \textit{purl} data field as shown in Figure~\ref{fig:compliance} and Table~\ref{tab:compliance-data}, which will lead to many zero purl consistency score that limits the sum tripe-factor score.
    \item \textbf{Relative ranking of tool pairs remains stable across threshold values:} The average Spearman's rank correlation coefficients for SBOM tool pairs within each language ecosystem are 0.9150 (C/C++), 0.8536 (Java), and 0.9247 (Python), all of which are statistically significant (p < 0.01). This high rank correlation indicates that the relative ordering of tool pairs remains consistent across different threshold values, confirming that our core finding regarding SBOM tool consistency limitations is robust and not an artifact of arbitrary threshold selection.
\end{itemize}

These results validate our choice of $\tau=2$ as the primary threshold for our main analysis. The stability of relative tool performance across thresholds further strengthens our conclusion that the observed consistency gaps reflect fundamental issues in SBOM tool implementation rather than methodological artifacts. This rigorous threshold analysis demonstrates the robustness of our evaluation framework and the validity of our core findings.

The design of our triple-factor matching algorithm is a direct response to the non-trivial challenge of identifier ambiguity within individual SBOMs. A simple name-based pairing is unreliable because a SBOM can list multiple, distinct packages that share the same name. To quantify the prevalence and nature of this challenge, we analyzed SBOMs in our dataset for intra-SBOM~duplications.

As shown in the column N of Table~\ref{tab:duplications}, the ambiguous packages with duplicated names commonly exist across SBOMs. The SBOMs produced by cdxgen show 21.66\%, 20.39\%, and 20.02\% ambiguous packages within C/C++, Java, and Python, respectively. 
This ambiguity can be partially alleviated by the triple-factor matching, as illustrated in the decreased duplication ratio within the ``NVP'' column. 
For instance, the gh-sbom tool generated a CycloneDX-format SBOM for the \texttt{asciidots}~\cite{asciidots} repository, it includes two components both named ``jekyll'' with version ``3.7.3''. 
However, they differ in their \textit{purl}: one is ``\texttt{pkg:gem/jekyll@3.7.3}'' and the other is ``\texttt{pkg:gem/jekyll@+3.7.3}''. 
Furthermore, the first component records a \textit{license} of ``MIT'', while the second reports no \textit{license} information. If the package matching method relies solely on \textit{package name} and \textit{version}, it may fail to correctly pair packages with their corresponding \textit{license} records. 
Our proposed triple-factor best-matching method mitigates such ambiguities and ensures deterministic matching results.

However, a few tools like syft produce SBOMs with perfect duplications, which represent an inherent ambiguity in the source SBOM data that no pairing algorithm can definitively resolve.
Our method cannot distinguish between such component, as shown in the most rows of SPDX results. 
In which, the proposed method can only provide a deterministic pairing result on the packages.

Overall, the design and settings of the evaluation module is effective and appropriate for the analysis on the adherence gap between SBOM standards and SBOM tools.

\begin{table*}[t]\small
    \vspace{-1em}
    \centering
    \caption{Duplicate ratio within the produced SBOMs by SBOM tools. The N, NV, NVP columns represent ``name'', ``name+version'', ``name+version+purl'', which are the duplicated fields within packages of SBOM.}
    \label{tab:duplications}
    \begin{tabular}{rrrrrrrrrrr} 
        \toprule
        & & \multicolumn{3}{c}{\textbf{C/C++}} & \multicolumn{3}{c}{\textbf{Java}} & \multicolumn{3}{c}{\textbf{Python}} \\
        \cmidrule[0.5pt](lr){3-5} \cmidrule[0.5pt](lr){6-8} \cmidrule[0.5pt](lr){9-11}
        & \textbf{Tool} & \textbf{N} & \textbf{NV} & \textbf{NVP} & \textbf{N} & \textbf{NV} & \textbf{NVP} & \textbf{N} & \textbf{NV} & \textbf{NVP} \\
        \midrule
        \multirow{5}{*}{\rotatebox{90}{\textbf{CycloneDX}}} & cdxgen & 21.66\% & 9.17\% & 7.51\%  & 20.39\% & 9.87\% & 4.43\%  & 20.02\% & 3.62\% & 1.76\%  \\
        & gh-sbom & 40.65\% & 18.71\% & 16.01\%  & 49.45\% & 12.71\% & 7.46\%  & 27.33\% & 7.98\% & 3.40\%  \\ 
        & ort & 0.72\% & 0.00\% & 0.00\%  & 3.71\% & 0.22\% & 0.22\%  & 0.36\% & 0.00\% & 0.00\%  \\
        & scancode & 5.48\% & 0.68\% & 0.10\%  & 3.47\% & 1.46\% & 0.09\%  & 1.12\% & 0.35\% & 0.00\%  \\ 
        & syft & 36.31\% & 32.23\% & 32.23\%  & 31.78\% & 28.86\% & 28.77\%  & 25.11\% & 23.90\% & 23.90\%  \\ 
        \midrule
        \multirow{4}{*}{\rotatebox{90}{\textbf{SPDX}}} & gh-sbom & 22.35\% & 8.92\% & 8.92\%  & 20.20\% & 3.41\% & 3.41\%  & 17.60\% & 1.18\% & 1.09\%  \\ 
        & ort & 7.62\% & 7.38\% & 7.38\%  & 16.03\% & 16.03\% & 16.03\%  & 24.28\% & 24.28\% & 24.28\%  \\ 
        & sbom-tool & 8.36\% & 0.19\% & 0.00\%  & 4.64\% & 0.27\% & 0.00\%  & 8.46\% & 0.17\% & 0.17\%  \\ 
        & syft & 36.31\% & 32.23\% & 32.23\%  & 31.78\% & 28.86\% & 28.77\%  & 25.11\% & 23.90\% & 23.90\%  \\
        \bottomrule
    \end{tabular}
\end{table*}

\subsubsection{Overall framework validation}

The empirical evaluations across \my's modules provide evidence of the framework's reliability for SBOM gap analysis. The SBOM generation module successfully processed 3,287 diverse repositories across multiple language ecosystems. The extraction module achieved a 94.2\% success rate, with failures primarily attributable to input issues rather than framework limitations. The evaluation module's ablation study on the threshold parameter $\tau$ showed consistent ranking of tool pairs across different threshold values (Spearman's $\rho>0.85$ for all languages), suggesting that our core findings regarding SBOM tool limitations are methodologically sound. The analysis on the ambiguity within SBOMs evidences the necessity and effectiveness of our designed triple-factor best-matching algorithm. These evaluations indicate that \my provides a reliable foundation for assessing SBOM standard adherence across tools, standards, and language ecosystems, with its modular design allowing for potential adaptations to specific evaluation needs.

\section{Discussions}

Our analysis reveals that the adherence gaps in SBOM generation are not merely technical flaws but symptoms of deeper, systemic friction within the software supply chain ecosystem. The observed inconsistencies between tools and standards point to a fundamental tension: the drive for rapid, automated tooling has outpaced the maturation of the standards they aim to implement. This section explores the underlying dynamics and proposes a path forward.

\textbf{Validation schema from SBOM standards:} The ambiguity in standards, particularly for fields like \textit{license} and \textit{version}, stems from an inherent paradox. On the one hand, overly prescriptive standards risk stifling innovation and failing to capture the nuanced realities of diverse package ecosystems, \textit{e.g.}, version schemes in Python vs. C++. On the other hand, the current flexibility breeds inconsistencies that undermine the SBOM's core promise of interoperability. 
Therefore, the challenge is not merely to ``clarify'' the standards, but to develop a more robust validation and conformance framework. This could involve machine-readable schemas with stricter definitions for core fields, coupled with optional, community-driven extensions for ecosystem-specific needs. This approach would transform standards from loose guidelines into enforceable contracts, ensuring a baseline of quality without sacrificing adaptability.

\textbf{Verification by SBOM tools:} Our findings indicate that most SBOM tools operate as passive metadata harvesters, relying on incomplete or untrustworthy local manifests. This limitation is a primary driver of inaccuracy. We advocate for a paradigm shift from passive generation to active verification. Tool developers should not only define their analysis scope transparently but also integrate mechanisms for cross-referencing against authoritative external sources, \textit{e.g.}, package registries, vulnerability databases, and provenance logs. Such a shift elevates the SBOM tool from a simple reporting utility to an active verification component within a DevSecOps pipeline, directly enhancing the integrity of the SBOM.

\textbf{Community shifts:} The downstream benefits of a high-quality SBOM, in streamlined vulnerability management and compliance automation, are clear. However, the incentives for tool developers and standards bodies are not perfectly aligned with producing the highest-fidelity artifacts. 
The community, armed with evaluation frameworks like ours, can correct this imbalance. This shifts the conversation from a technical recommendation to an economic and strategic one, empowering the SBOM community to drive meaningful change.

\section{Threats to Validity}

This section discusses potential threats to the findings and measures taken to mitigate their impacts.

\textbf{Internal threats:}
To enhance the generalizability of the evaluation findings, our evaluations primarily assess the adherence of SBOM tools to widely recognized requirements, specifically the mandatory and $NTIA^+$ fields stipulated by SBOM standards and governmental guidelines.
However, this focus on common, standardized fields may not comprehensively cover all custom scenarios or specific data field requirements of individual SBOM users.
Moreover, we adopted the officially recommended parameters of each SBOM tool from its documentation to ensure the effectiveness of evaluations, which can also introduce bias from the scenario of users. To address this, our evaluation framework, \my, is designed with a flexible pipeline that aligns with SBOM standards and is capable of handling custom data fields and the parameters of SBOM tools based on specific user needs. 
Therefore, users can adapt and utilize \my to assess SBOM tools against their own unique scenarios and requirements, enabling them to identify the most suitable tool for their particular use case.

\textbf{External threats:}
Our evaluation's scope presents certain limitations. 
The primary dataset is confined to GitHub source code repositories with the number of stars as a selecting proxy (excluding binaries/images that might alter tool analysis), and our accuracy assessment relies on a Python-only ground truth dataset, $D_{gt}$, which restricts the quantitative generalization of these specific accuracy metrics to other language ecosystems. We acknowledge that focusing on a single language is a limitation, as different ecosystems exhibit unique dependency management practices that can impact tool accuracy. However, the fundamental methodology we introduce for constructing $D_{gt}$, a multi-stage verification process combining automated extraction with human validation, is language-agnostic. Extending our dataset to other languages like Java or Golang is primarily an engineering challenge that involves scaling up our verification pipeline to accommodate different manifest formats and repository structures. This constitutes a significant but straightforward engineering effort, rather than a change to the core scientific contribution of our work, which is the rigorous evaluation methodology itself. Nevertheless, the focus on SBOM tool adherence to standards remains broadly relevant.
The fundamental challenges causing the observed adherence gaps appear to be common across diverse software ecosystems, as indicated by the compliance and consistency evaluations. 
This observation can support the generalization of qualitative findings from the accuracy evaluations.
Future work aims to address these limitations by evaluating SBOM tools on a broader range of software formats and developing multilingual ground truth datasets for accuracy assessments.

\section{Related Works}

\subsection{Evaluation on SBOMs}

The rapid growth in SBOM adoption in recent years has spurred a corresponding surge in academic and industry research focused on its capabilities and limitations~\cite{10336262, 10305922, tobar2025software, nocera2025adoption}.
Several prior studies have evaluated the quality of SBOMs, aiming to understand their current state and identify areas for improvement~\cite{fossa-blog, sbombenchmark, 10315783, xiao2025jbomaudit}. 
However, they often analyze individual SBOM instances without a comprehensive comparison across a diverse corpus or detailed scrutiny against standard requirements. The \cite{sbombenchmark} analyzes a single SBOM uploaded to its website and calculates the existing data fields. The JbomAudit~\cite{xiao2025jbomaudit} assesses the completeness and accuracy of packages identified by the Jbom tool within given JAR files, comparing with the manually mapped dependency graph.
Some recent works have analyzed the accuracy of SBOM tools~\cite{cofano2024sbom, 10.1145/3605098.3635927, halbritter2024accuracy}. The~\cite{cofano2024sbom} analyzes the SBOM generation tools in the Python ecosystem by employing the tools to produce SBOMs on their synthetic Python projects. The~\cite{10.1145/3605098.3635927} analyzes the SBOM tools within the npm ecosystem on 50 GitHub projects. Similarly, the~\cite{halbritter2024accuracy} evaluated the accuracy of several SBOM tools in the area of web application tools on synthetic projects.  However, their focuses on package-level analysis limit the scope of results and cannot reflect the accuracy of detailed information within packages.
The study by \cite{yu2024correctness} utilized differential comparisons of SBOM tools to evaluate the accuracy of SBOM tools. 
Nevertheless, these studies did not primarily focus on adherence to SBOM standards or the complexities inherent in real-world software.

To address these limitations, this paper investigates the adherence gap between SBOM standards and tools in large-scale real-world repositories, focusing on key data quality attributes grounded in the ISO/SEC 25012 model. As detailed in Table~\ref{tab:related_works}, our work distinguishes itself by conducting a granular analysis within packages and rigorously using the SBOM standards as the reference baseline. This enables a comprehensive understanding of these gaps and their practical implications.

\begin{table}[t]\small
  \vspace{-1em}
  \caption{Differences between existing works. The ``scale'' is the number of projects or files the paper analyzed.}
  \label{tab:related_works}
  \centering
  \resizebox{\linewidth}{!}{
  \begin{tabular}{ccccccc} 
    \toprule
     \textbf{Studies} & \textbf{Real-world} & \textbf{Scale} & \makecell{\textbf{Multiple}\\ \textbf{programming language}} & \makecell{\textbf{Package-level} \\ \textbf{analysis}} & \makecell{\textbf{Field-level} \\\textbf{analysis}} & \makecell{\textbf{Adherence to} \\ \textbf{SBOM standards}} \\
    \midrule
    \textbf{Cofano \textit{et al.}~\cite{cofano2024sbom}}  & \ding{56} & 10 & \ding{56} & \ding{51} & \ding{56} & \ding{56}\\
    \textbf{Rabbi \textit{et al.}~\cite{10.1145/3605098.3635927}}  &  Github Projects & 50 & \ding{56} & \ding{51} & \ding{56} & \ding{56} \\
    \textbf{Halbritter \textit{et al.}~\cite{halbritter2024accuracy}}  & \ding{56} & 4 & \ding{51} & \ding{51} & \ding{56} & \ding{56} \\
    \textbf{Xiao \textit{et al.}~\cite{xiao2025jbomaudit}} & Java JARs & 25,882 & \ding{56} & \ding{51} & \ding{56} & \ding{56} \\
    \textbf{Yu \textit{et al.}~\cite{yu2024correctness}}  &  Github Projects & 7,876 & \ding{51} & \ding{51} & \ding{56} & \ding{56} \\
    \textbf{Ours}  & Github Projects & 3,287 & \ding{51} & \ding{51} & \ding{51} & \ding{51} \\
    \bottomrule
  \end{tabular}
  }
\end{table}

\subsection{Software supply chain security}

The software supply chain substantially improves development efficiency in modern software~\cite{synkshift, ntiaframing}. However, it also expanded the attack surface, exposing software to supply chain-centric attacks~\cite{williams2025research, shen2025understanding, ladisa2023sok, liang2023needle, 11215664, cybersecuritygraph2024}. Attacks like Log4J~\cite{log4j}, SolarWinds~\cite{alkhadra2021solar}, XZ utils~\cite{przymus2025wolves}, have threatened thousands of software, making the management of the software supply chain of vital importance.

Various approaches have been proposed to alleviate the risks within the software supply chain~\cite{ohm2023sok,9985180,vu2020towards,zheng2024towards, 11275815, LibSleuth2025}. Software Composition Analysis (SCA) and SBOM are important approaches that can manage the dependencies and their detailed information for transparency~\cite{sok,20252604,intelligentvehicles2024}. The SCA focuses on identifying and managing third-party components, approaches like CENTRIS~\cite{woo2021centris}, CCScanner~\cite{tang2022towards}, TPLite~\cite{jiang2023third},  CNEPS~\cite{na2024cneps}, BinaryAI~\cite{jiang2024binaryai}, VISION~\cite{wu2024vision}, TIVER~\cite{choi2025tiver} kept evolving to get accurate dependencies of software. These tools are engineered to parse complex dependency graphs and report the found dependent packages.
Modern SCA tools increasingly generate SBOMs as standardized output, positioning SBOMs as the data layer that underpins advanced vulnerability management and risk assessment~\cite{cdxgen, gh-sbom, ort, sbom-tool, scancode,syft}. Researchers investigate the dependencies of the language ecosystems to find potential errors~\cite{latendresse2022not}. Some studies dive into the version management and vulnerability propagation to further understand the security landscape of the software supply chain~\cite{li2023large,hu2024empirical,liu2022demystifying}. These studies broaden the downstream applications of both SCA and SBOM tools.

While extensive research has advanced SCA tools and security frameworks, the quality and accuracy of the SBOM artifact itself remains unchecked. 
By systematically measuring the adherence of SBOM tools to standards, our work provides a crucial empirical basis to understand and improve the trustworthiness of SBOMs, thereby strengthening the security ecosystem that relies on them.

\section{Conclusions}

This paper presents a large-scale, two-stage empirical analysis of the adherence gaps between SBOM standards and tools using our extensible framework, \my.
Our evaluation, comprising a baseline benchmark and a one-year longitudinal follow-up, encompasses 55,444 SBOMs generated by six leading tools from 3,287 real-world repositories.
Although SBOM tools achieve basic interoperability with SBOM standards, this study observes significant and persistent gaps, including: inadequate adherence to policy requirements; poor inter-tool consistency in package detection (under 13\% on average); high \textit{longitudinal} inconsistency, where tools conflict with their own prior versions; and accuracy rates of no more than 20\% for package license information.
Our two-stage analysis demonstrates these gaps are systemic, persistent, and volatile, attributed to vague standard constraints and the unpredictable nature of tool evolution.
This paper also discusses solutions for addressing these gaps.



\section*{Acknowledgment}

This paper is supported by the National Key R\&D Program of China (2024YFB4506200) and the YuanTu Large Research Infrastructure. We are deeply grateful to the anonymous reviewers and the editors for their insightful comments, which have significantly improved this paper.

\bibliographystyle{ACM-Reference-Format}
\bibliography{sbom}

\begin{appendices}

\section{Longitudinal Follow-up Study: Full Results}
This appendix provides the detailed results from our one-year longitudinal follow-up study conducted in October 2025. The experimental setup and dataset remained identical to the 2024 baseline analysis, with the only change being the updated versions of the six evaluated SBOM tools\footnote{~The versions of the SBOM tools in the follow-up study are: cdxgen@v11.8.0, gh-sbom@v0.1.1, ort@v69.0.0, syft@v1.33.0, sbom-tool@v4.1.1, and scancode@v32.4.1.}. The target SBOM standards kept as SPDX 2.3 and CycloneDX 1.5. This follow-up stage yielded 27,649 new SBOMs, of which 25,330 (91.6\%) were successfully processed by our framework.

\subsection{Compliance results (follow-up)}

The compliance results from our follow-up analysis corroborate the findings of our baseline, demonstrating that the identified compliance gaps are persistent challenges within the SBOM ecosystem. The detailed results are presented in Table~\ref{app-tab:compliance-data}. Our key observations are as follows:

\begin{table}[b]\small
    \centering
    \caption{Detail compliance evaluation results across languages for the evolved SBOM tools. The ``Mandatory'' column includes results for all mandatory fields of each standard, while the remaining data fields are from $NTIA^+$ (the \textit{creator} field is overlapped in ``mandatory''). All the results are reported in percentages.}
    \label{app-tab:compliance-data}
    \resizebox{\linewidth}{!}{
    \begin{tabular}{rrrrrrrrrr}
    \toprule
        & \textbf{Tool} & \textbf{Mandatory} & \textbf{name} & \textbf{supplier} & \textbf{purl} & \textbf{license} & \textbf{version} & \textbf{copyright} & \textbf{relationship} \\ \midrule
        \multirow{6}{*}{\rotatebox{90}{\textbf{C/C++}}} & syft & 100.00 & 100.00 & 33.41 & 45.76 & 1.84 & 51.29 & 0.00 & 100.00 \\ 
        & gh-sbom & 100.00 & 100.00 & 0.00 & 100.00 & 28.26 & 96.70 & 10.59 & 100.00 \\ 
        & ort & 100.00 & 100.00 & 3.83 & 3.30 & 83.63 & 78.07 & 86.00 & 5.13 \\ 
        & sbom-tool & 100.00 & 100.00 & 88.69 & 99.97 & 5.73 & 99.97 & 0.00 & 100.00 \\ 
        & scancode$^\dag$ & 100.00 & 29.61 & 11.42 & 29.61 & 29.61 & 17.58 & 1.11 & 100.00 \\ 
        & cdxgen$^\dag$ & 100.00 & 67.06 & 0.00 & 67.06 & 3.93 & 67.06 & 0.00 & 100.00 \\ 
        \midrule
        \multirow{6}{*}{\rotatebox{90}{\textbf{Java}}} & syft & 100.00 & 100.00 & 7.60 & 65.96 & 4.60 & 66.49 & 0.00 & 100.00 \\ 
        & gh-sbom & 100.00 & 100.00 & 0.00 & 100.00 & 33.79 & 90.41 & 13.58 & 100.00 \\ 
        & ort & 100.00 & 100.00 & 6.51 & 5.93 & 85.23 & 17.46 & 72.01 & 7.55 \\ 
        & sbom-tool & 100.00 & 100.00 & 94.60 & 100.00 & 4.44 & 100.00 & 0.00 & 100.00 \\ 
        & scancode$^\dag$ & 100.00 & 33.42 & 3.10 & 33.42 & 33.42 & 32.01 & 0.01 & 100.00 \\ 
        & cdxgen$^\dag$ & 100.00 & 92.90 & 0.00 & 92.90 & 0.85 & 92.90 & 0.00 & 100.00 \\ 
        \midrule
        \multirow{6}{*}{\rotatebox{90}{\textbf{Python}}} & syft & 100.00 & 100.00 & 24.26 & 50.49 & 9.21 & 51.86 & 0.00 & 100.00 \\ 
        & gh-sbom & 100.00 & 100.00 & 0.00 & 100.00 & 22.14 & 78.76 & 19.22 & 100.00 \\ 
        & ort & 100.00 & 100.00 & 14.61 & 9.64 & 74.05 & 55.97 & 70.97 & 11.53 \\ 
        & sbom-tool & 100.00 & 100.00 & 93.91 & 100.00 & 4.03 & 100.00 & 0.00 & 100.00 \\ 
        & scancode$^\dag$ & 100.00 & 56.22 & 43.33 & 56.22 & 56.22 & 34.55 & 0.40 & 100.00 \\ 
        & cdxgen$^\dag$ & 100.00 & 70.50 & 0.00 & 70.50 & 32.45 & 70.50 & 0.00 & 100.00 \\ 
        \bottomrule
    \end{tabular}
    }
    
    \raggedright\footnotesize
    ~$^\dag$ Tools that report results in CycloneDX standard, the unmarked tools are reported in SPDX standards.
    \vspace{-2em}
\end{table}

\begin{itemize}
    \item \textbf{Persistence of systemic deficiencies:} The most significant finding is the continued, systemic failure of tools to provide complete data for key policy-mandated ($NTIA+$) fields. Mirroring our baseline results, fields such as \textit{supplier} and \textit{copyright} remain sparsely populated across most tools. For example, \textit{sbom-tool}, while generally compliant, still exhibits near-zero compliance for the \textit{copyright} field and only 5.73\% for \textit{license} in C/C++ projects. This indicates that these are deep-rooted implementation challenges, not transient bugs that were fixed over the year.
    \item \textbf{Stable adherence to foundational requirements:} Consistent with the baseline, all evolved tool versions demonstrate 100\% compliance with the basic mandatory fields defined by the SPDX and CycloneDX standards, \textit{i.e.}, \textit{spec version}, \textit{bom format}, \textit{etc.} as defined in the Table~\ref{tab:compliance}. This reinforces our conclusion that while tools can produce structurally valid SBOMs, this offers no guarantee of their semantic completeness or practical utility.
    \item \textbf{Consistent language-dependent patterns:} The performance dichotomy between language-specific and language-agnostic fields also persists. Data fields that can be derived from package manager metadata, \textit{e.g.}, \textit{package name}, \textit{version}, \textit{purl}, continue to show variable compliance depending on the language ecosystem. Conversely, language-agnostic fields like \textit{license} and \textit{copyright} exhibit consistently poor compliance across all languages.
\end{itemize}

In summary, the longitudinal compliance data provides strong evidence that the gaps identified in our baseline study are not isolated incidents but are indicative of persistent, systemic issues in the SBOM toolchain. The ecosystem has shown only marginal improvement in addressing these fundamental compliance challenges over a one-year period.

\subsection{Consistency results (follow-up)}
\label{app:consistency}

\begin{table*}[b]\small 
    \centering
    \caption{Consistency evaluation results on the evolved versions of SBOM tools. The pkg., ver., and lic. represents the package, version, and license. The package consistency is calculated by Formula~(\ref{eqa:consistency-pkg}), and the consistency of the other data fields is calculated with Formula~(\ref{eqa:consistency-field}). Reporting results in percentages.}
    \label{app-tab:consistency}
    \resizebox{\linewidth}{!}{
    \begin{tabular}{rrrrrrrrrrrrrr} 
        \toprule
        \multicolumn{2}{c}{} & \multicolumn{4}{c}{\textbf{C/C++}} & \multicolumn{4}{c}{\textbf{Java}} & \multicolumn{4}{c}{\textbf{Python}} \\
        \cmidrule[0.5pt](lr){3-6} \cmidrule[0.5pt](lr){7-10} \cmidrule[0.5pt](lr){11-14}
        & \textbf{Tool pair} & \textbf{pkg.} & \textbf{ver.} & \textbf{purl} & \textbf{lic.}  & \textbf{pkg.} & \textbf{ver.} & \textbf{purl} & \textbf{lic.} & \textbf{pkg.} & \textbf{ver.} & \textbf{purl} & \textbf{lic.}\\
        \midrule
        \multirow{10}{*}{\rotatebox{90}{\textbf{CycloneDX}}} & syft$\leftrightarrow$gh-sbom & 5.30 & 15.08 & 15.82 & 0.77 & 31.66 & 60.51 & 70.64 & 0.77 & 20.18 & 42.98 & 44.59 & 0.63 \\ 
        ~ & syft$\leftrightarrow$ort & 0.54 & 1.87 & 1.68 & 0.10 & 1.95 & 6.14 & 6.15 & 0.11 & 1.44 & 4.45 & 4.16 & 0.00 \\ 
        ~ & syft$\leftrightarrow$scancode & 0.99 & 8.18 & 8.96 & 1.73 & 4.56 & 30.09 & 29.87 & 1.51 & 4.63 & 24.94 & 25.54 & 8.18 \\ 
        ~ & syft$\leftrightarrow$cdxgen & 7.27 & 17.88 & 17.38 & 0.50 & 34.01 & 85.28 & 71.70 & 0.29 & 15.90 & 33.33 & 33.32 & 0.40 \\ 
        ~ & gh-sbom$\leftrightarrow$ort & 2.67 & 5.89 & 5.19 & 0.95 & 11.52 & 32.75 & 33.70 & 17.11 & 4.66 & 10.22 & 10.38 & 3.58 \\ 
        ~ & gh-sbom$\leftrightarrow$scancode & 0.33 & 2.48 & 2.62 & 0.55 & 2.18 & 23.66 & 24.65 & 1.39 & 0.15 & 1.19 & 2.17 & 0.47 \\ 
        ~ & gh-sbom$\leftrightarrow$cdxgen & 44.27 & 67.13 & 81.63 & 2.40 & 31.49 & 80.71 & 77.53 & 0.22 & 47.01 & 75.35 & 78.08 & 11.10 \\ 
        ~ & ort$\leftrightarrow$scancode & 0.05 & 0.37 & 0.33 & 0.22 & 0.03 & 0.43 & 0.46 & 0.13 & 0.00 & 0.00 & 0.00 & 0.00 \\ 
        ~ & ort$\leftrightarrow$cdxgen & 1.88 & 3.56 & 3.46 & 0.41 & 4.06 & 5.65 & 4.72 & 0.04 & 4.59 & 7.33 & 7.58 & 2.15 \\ 
        ~ & scancode$\leftrightarrow$cdxgen & 0.48 & 4.95 & 4.93 & 0.79 & 1.08 & 12.64 & 10.83 & 0.11 & 0.24 & 2.70 & 2.93 & 0.61 \\ 
        \midrule
        \multirow{6}{*}{\rotatebox{90}{\textbf{SPDX}}} & syft$\leftrightarrow$gh-sbom & 30.78 & 41.38 & 49.19 & 0.00 & 7.97 & 20.06 & 23.62 & 0.00 & 25.67 & 42.99 & 47.18 & 0.00 \\ 
        ~ & syft$\leftrightarrow$ort & 0.90 & 4.37 & 1.77 & 0.72 & 1.74 & 8.58 & 4.67 & 0.51 & 2.49 & 10.27 & 4.30 & 2.34 \\ 
        ~ & syft$\leftrightarrow$sbom-tool & 5.66 & 10.10 & 9.95 & 0.00 & 2.64 & 5.66 & 5.62 & 0.00 & 3.40 & 7.01 & 7.02 & 0.00 \\ 
        ~ & gh-sbom$\leftrightarrow$ort & 0.47 & 2.23 & 2.31 & 0.00 & 0.06 & 0.40 & 0.38 & 0.00 & 0.75 & 4.90 & 5.06 & 0.00 \\ 
        ~ & gh-sbom$\leftrightarrow$sbom-tool & 4.71 & 9.60 & 9.96 & 4.46 & 3.80 & 6.44 & 6.75 & 4.94 & 2.89 & 5.72 & 6.01 & 3.36 \\ 
        ~ & ort$\leftrightarrow$sbom-tool & 1.94 & 6.13 & 1.94 & 0.00 & 0.43 & 1.71 & 0.23 & 0.00 & 0.74 & 3.17 & 0.40 & 0.00 \\ 
        \midrule
        ~ & \textbf{Average$^\dag$} & \textbf{6.77} & \textbf{12.58} & \textbf{13.57} & \textbf{0.85} & \textbf{8.70} & \textbf{23.79} & \textbf{23.22} & \textbf{1.70} & \textbf{8.42} & \textbf{17.28} & \textbf{17.42} & \textbf{2.05} \\ 
        \bottomrule
    \end{tabular}
    }
    
    \raggedright\footnotesize
    ~$^\dag$ Average consistency score of a data field across all tool pairs for each language.
    \vspace{-2em}
\end{table*}

This section presents the inter-tool consistency results from our longitudinal follow-up study. The primary objective is to assess whether the critically low consistency observed in our baseline has improved over a one-year development cycle, or if it represents a persistent, systemic challenge. The detailed results are presented in Table~\ref{app-tab:consistency}.
Our analysis confirms the persistence of the consistency gap and reinforces our baseline conclusions with the following key observations:

\begin{itemize}
    \item \textbf{Persistence of critically low inter-tool consistency:} The overall inter-tool consistency remains exceptionally low, showing no meaningful improvement from our 2024 baseline. In 2025, the average inter-tool consistency for package detection was 6.77\%, 8.70\% and 8.42\% for C/C++, Java and Python respectively. These consistency scores are not statistically different from the baseline results (7.84\%, 12.77\%, and 10.09\% respectively), which strongly indicates that the root causes of inconsistency, such as ambiguous standard interpretations and divergent tool implementations, have not been resolved at an ecosystem~level.
    \item \textbf{Replication of inconsistency patterns:} The patterns of inconsistency observed in the follow-up mirror those from our baseline. Data fields that are difficult to parse or require external lookups, such as \textit{license}, continue to exhibit near-zero consistency across most tool pairs. In contrast, fields that can be more easily derived from package manager metadata, like \textit{version} and \textit{purl}, show comparatively higher consistency. This replication of patterns reinforces that the inconsistency is systemic and tied to specific, unsolved technical challenges.
    \item \textbf{Volatility and unpredictability of the ecosystem:} While the ecosystem-wide average shows stagnation, the performance of individual tool pairs is highly volatile. For example, the consistency between gh-sbom and cdxgen in the Python package detection may have marginally improved from 32.83\% to 47.01\%, while the pairing of syft and cdxgen in the Java package detection regressed from 63.90\% to 34.01\%. This unpredictability is a significant finding in itself, suggesting that practitioners cannot rely on a linear or guaranteed improvement trajectory when selecting or updating their tools. The ecosystem's evolution is not uniform, making consistent SBOM generation a moving target.
\end{itemize}

In conclusion, the longitudinal consistency data provides compelling evidence that the interchangeability of SBOMs remains a major, unsolved challenge. The gaps we identified are persistent features of the current SBOM toolchain, validating the conclusions drawn from our baseline.

\subsection{Accuracy results (follow-up)}
\label{app:accuracy}

This section presents the accuracy results of the follow-up, evaluated against our Python ground truth dataset $D_{gt}$. The goal is to determine whether the accuracy gaps identified in our baseline are persistent. The detailed results are in Table~\ref{app-tab:acc}.
The follow-up data reveals that accuracy remains a critical, systemic failure for the SBOM ecosystem. Furthermore, it highlights the potential volatility of tool performance, where ``improvements'' are not guaranteed and severe regressions can occur.

\begin{table}[t]\small
    \vspace{-1.5em}
    \centering
    \caption{Accuracy evaluation results on the evolved SBOM tools. ``Count'' means the evaluated valid SBOMs.}
    \label{app-tab:acc}
    \begin{minipage}{0.75\linewidth}
    \centering
    \begin{tabular}{rrrrrrr}
        \toprule
        \textbf{Tool} & \textbf{Count} & \textbf{Precision$^\dag$} & \textbf{Recall$^\dag$} & \textbf{supplier} & \textbf{license} & \textbf{version} \\ 
        \midrule
        syft & 100 & 11.47\% & 13.89\% & 0.06\% & 0.00\% & 18.74\% \\ 
        gh-sbom & 60 & 71.87\% & 61.56\% & 0.00\% & 10.91\% & 64.77\% \\ 
        ort & 84 & 6.31\% & 9.98\% & 0.00\% & 7.97\% & 2.68\% \\ 
        scancode & 100 & 1.00\% & 0.04\% & 1.00\% & 1.00\% & 0.00\% \\ 
        cdxgen & 99 & 35.70\% & 33.26\% & 0.00\% & 22.04\% & 21.84\% \\ 
        sbom-tool$^\ddag$ & 100 & 1.05\% & 0.17\% & 0.42\% & 1.00\% & 0.81\% \\ 		
        \bottomrule
    \end{tabular}
    
    \raggedright\footnotesize
    ~$^\dag$ The precision and recall are calculated on package detection as in \S \ref{sec:acc-design}.\\
    ~$^\ddag$ The sbom-tool is in SPDX, the other tools are in CycloneDX.
    \end{minipage}
\end{table}

\begin{itemize}
    \item \textbf{Persistence of mediocre to poor accuracy:} The overall accuracy landscape in the one-year follow-up mirrors the poor performance of the baseline. While gh-sbom (71.87\% Precision / 61.56\% Recall) and cdxgen (35.70\% Precision / 33.26\% Recall) retain a moderate capability for package detection, the ecosystem as a whole has not solved this problem.
    \item \textbf{Volatility and performance regression:} The evaluation results show the volatility of tool accuracy. While gh-sbom's performance remained relatively stable, sbom-tool's package detection capability collapsed dramatically, with its Precision/Recall dropping from 14.50\%/24.38\% in the baseline to a near-zero 1.05\%/0.17\% in the evolved version. This regression demonstrates that tool evolution may \textit{degrade} reliability.
    \item \textbf{Systemic failure in field-level accuracy:} Consistent with our baseline, all tools, including those with moderate package detection, continue to fail at providing accurate field-level details. Accuracy for the \textit{license} field remains below 22.04\% (achieved by cdxgen).
\end{itemize}

In summary, the longitudinal accuracy analysis provides compelling evidence that the accuracy gaps are deep-seated and persistent. The ecosystem has not seen meaningful improvement after one year, and the high volatility, evidenced by significant performance regressions in some tools, poses a severe risk to any organization relying on these tools for accurate software supply chain~data.
\end{appendices}

\end{document}